\documentclass[fleqn,usenatbib]{mnras}

\usepackage{newtxtext,newtxmath}
\usepackage[T1]{fontenc}

\DeclareRobustCommand{\VAN}[3]{#2}
\let\VANthebibliography\thebibliography
\def\thebibliography{\DeclareRobustCommand{\VAN}[3]{##3}\VANthebibliography}

\usepackage{graphicx}
\usepackage{float}	% Including figure files
\usepackage{amsmath}	% Advanced maths commands
\usepackage{multirow,bigdelim}

\defcitealias{2024MNRAS.528.7564B}{B24}
\newcommand{\ballard}{\citetalias{2024MNRAS.528.7564B}} 

\defcitealias{2024arXiv240811090M}{M24}
\newcommand{\minor}{\citetalias{2024arXiv240811090M}}

\defcitealias{2024MNRAS.52710480N}{N24}
\newcommand{\nightingale}{\citetalias{2024MNRAS.52710480N}}

\defcitealias{2024arXiv240712910D}{D24}
\newcommand{\despali}{\citetalias{2024arXiv240712910D}}

\usepackage{colortbl}
\usepackage{overpic} 
\usepackage{xcolor}

\usepackage{tikz}
\usetikzlibrary{shapes.geometric, arrows, backgrounds, positioning, fit}

\newcommand{\DeflectionAngle}{ \vec \alpha(\vec x) }

\newcommand{\jackpot}{J0946+1006 } 

\title[A subhalo undergoing gravothermal collapse?]{The overconcentrated dark halo in the strong lens SDSS~J0946+1006 is a subhalo: evidence for self interacting dark matter?}

\author[W. J. R. Enzi et al.]{
Wolfgang J. R. Enzi,$^{1,}$\thanks{E-mail: wolfgang.enzi@port.ac.uk}
Coleman M. Krawczyk$^1$,
Daniel J. Ballard$^{1,2}$,
and Thomas E. Collett$^1$
\\
$^1$ Institute of Cosmology and Gravitation (ICG), University of Portsmouth, Burnaby Rd, Portsmouth PO1 3FX, UK\\
$^2$ Sydney Institute for Astronomy, School of Physics, University of Sydney, NSW 2006, Australia
}

\date{Accepted XXX. Received YYY; in original form ZZZ}

\pubyear{2024}

\begin{document}
\label{firstpage}
\pagerange{\pageref{firstpage}--\pageref{lastpage}}
\maketitle

\begin{abstract}
The nature of dark matter is poorly constrained
on subgalactic scales. Alternative models to cold dark matter, such as warm dark matter or self-interacting dark matter, could produce very different dark haloes on these scales.
One of the few known dark haloes smaller than a galaxy was discovered in the triple source plane strong lens system J0946+1006. Previous studies have found that this structure is much more concentrated than expected in  $\Lambda$CDM, but have assumed the dark halo is at the same redshift as the main deflector ($z_{\rm main}=0.222$).
In this paper, we fit for the redshift of this dark halo. We reconstruct the first two sources in the system using a forward modelling approach, allowing for additional complexity from multipole perturbations. We find that the perturber redshift is $z_{\rm halo} = {0.207}^{+0.019}_{-0.019}$, and lower bounds on the evidence strongly prefer a subhalo over a line-of-sight structure. Whilst modelling both background sources does not improve constraints on the redshift of the subhalo,  it breaks important degeneracies affecting the reconstruction of multipole perturbations. 
We find that the subhalo is a more than $5\sigma$ outlier from the $\Lambda$CDM  $v_{\rm max}$--$r_{\rm max}$ relation and has a steep profile with an average slope of $\gamma_{\rm 2D} = {-1.81}^{+0.15}_{-0.11}$ for radii between $0.75-1.25$ kpc. This steep slope might indicate dark matter self-interactions causing the subhalo to undergo gravothermal collapse; such collapsed haloes are expected to have $\gamma_{\rm 2D} \approx -2$.
\end{abstract} % 240 words


%%%%%%%%%%%%%%%%%%%%%%%%%%%%%%%%%%%%%%%%%%%%%%%%%%

%%%%%%%%%%%%%%%%% BODY OF PAPER %%%%%%%%%%%%%%%%%%

\section{Introduction}

While $\Lambda$CDM makes successful predictions about the large-scale structure of the Universe, no particle candidate for dark
matter (DM) has so far been detected in the laboratory. Although it has been studied extensively, cold dark matter (CDM) is, therefore, merely a placeholder for a more fundamental theory of dark matter. It is crucial for our understanding of astrophysics and cosmology that we place stronger constraints on its nature. The standard $\Lambda$CDM model predicts the formation of dark matter haloes that closely follow NFW mass profiles \citep[][]{1997ApJ...490..493N}. State-of-the-art hydrodynamical simulations \citep[e.g.][]{2015MNRAS.446..521S,2019ComAC...6....2N,2020Natur.585...39W,2023MNRAS.524.2556H} indicate clear relations among the properties of these haloes, e.g. between their concentration and mass  \citep[see e.g.][]{2024arXiv240901758S} or between the maximum circular velocity and the corresponding radius at which this velocity is found \citep[see e.g.][]{2023MNRAS.518..157M}. 

However, the properties of dark matter that affect galactic and subgalactic scales are still not fully constrained. Alternatives to CDM might provide a better description of structure formation on these scales. For example, Warm dark matter (WDM) would be preferred if there is a suppressed number of small mass dark matter haloes \citep[see e.g.][]{2001ApJ...556...93B,2014MNRAS.439..300L,2020ApJ...897..147L}. Models such as fuzzy dark matter \citep[FDM, see e.g.][]{2000PhRvL..85.1158H,2017PhRvD..95d3541H} or self-interacting dark matter \citep[SIDM, see e.g.][]{2012MNRAS.423.3740V,2016MNRAS.460.1399V,2016PhRvD..93l3527C,2019MNRAS.484.4563D,2024JCAP...02..032Y} could further explain observed cores within dwarf galaxies.

SIDM is particularly interesting, as it can generate a variety of haloes with both cores and cusps. Recently, velocity-dependent SIDM became more popular since these models can create cores in small galaxies while allowing Milky way-mass haloes to remain non-spherical and therefore consistent with observations \citep[][]{2012MNRAS.423.3740V,2013MNRAS.431L..20Z}. A SIDM halo is initially cored because the self-interaction redistributes energy and momentum mostly in the centre where most interactions occur. The core then expands as heat flows inwards from the outskirts of the halo until it becomes isothermal. This process leads to halo profiles that are less cuspy in their centre than their CDM counterparts \citep[see e.g.][]{2000PhRvL..84.3760S,2024MNRAS.529.4611S,2024JCAP...02..032Y}. The expansion continues until a strong enough negative energy gradient is established and the random motion of particles in the core is no longer sufficient to support its own gravity. This leads eventually to gravothermal collapse, resulting in haloes with more cuspy density profiles than their CDM cousins \citep[see e.g.][]{2021MNRAS.505.5327T,2024JCAP...02..032Y}. Core collapse 
 behaves similar to the gravothermal catastrophe found in globular clusters \citep[][]{1980MNRAS.191..483L}.

While these alternative models make predictions that differ from CDM, the presence of baryons can also resculpt haloes without the need for exotic dark matter models \citep[see e.g.][who provide a discussion specifically in the context of gravitational lensing]{2023arXiv230611781V}.  Baryonic physics will further affect dark halo concentrations \citep[see e.g.][]{2024MNRAS.52711996H} and the expected luminosity of the galaxies that they host \citep[see e.g.][]{2024arXiv240712910D}.

Strong gravitational lensing has been employed extensively to study WDM and sterile neutrinos \citep[see e.g.][]{2018MNRAS.481.3661V,2019MNRAS.485.2179R,2020MNRAS.491.6077G,2021MNRAS.506.5848E,2021ApJ...917....7N}, SIDM \citep[see e.g.][]{2022MNRAS.516.4543D,2023PhRvD.107j3008G,2024ApJ...965L..19K}, and FDM \citep[see e.g.][]{2023MNRAS.524L..84P,2023NatAs...7..736A}. 
It can be used to detect and constrain the profiles of dark haloes from their localised lensing effect on multiple images, therefore allowing us to draw conclusions on dark matter microphysics and baryonic physics. However, so far only a small number of subhalos (i.e. haloes that are hosted by more massive haloes) has been detected this way \citep[less than five, ][]{2010MNRAS.408.1969V,2012Natur.481..341V,2014MNRAS.442.2434N,2016ApJ...823...37H}. 

A particularly interesting dark perturber has been detected by \citet{2010MNRAS.408.1969V} in the lens system is J0946+1006 \citep[which is part of the SLACS sample of lens systems, see][]{2008ApJ...677.1046G}.
Previous studies of this dark halo have found it to be unusually overconcentrated \citep[][]{2021MNRAS.507.1662M}. 
Recently, \citet[][referred to as \despali~ from here onwards]{2024arXiv240712910D} has further shown that it might further be an outlier in terms of its luminosity. \citet[][referred to as \minor~ from here onwards]{2024arXiv240811090M} found that supersampling plays a significant role in the detection significance. In agreement with \citet[][referred to as \ballard~ from here onwards]{2024MNRAS.528.7564B}, they also find that the inclusion of the second source breaks several degeneracies in the lens model. Nonetheless, the concentration parameter remains in tension with $\Lambda$CDM.

 The contributed lensing effect from line-of-sight haloes can be more important than the one from subhaloes, depending on the geometry of a lens system \citep[see, e.g.][]{2018MNRAS.475.5424D}. If the dark halo found in \jackpot is a line-of-sight halo, we would expect that its redshift has important implications for the observed arcs since compound lensing can lead to qualitatively different images than single plane lensing \citep[][]{2016MNRAS.456.2210C}. Only some of these effects can be absorbed by other parameters in the lens model, e.g. the mass distribution of the main deflector or the source light distribution
\citep[see e.g.][]{2017MNRAS.468.1426L,2022MNRAS.510.2464A}. We, therefore, expect to constrain the redshift of this dark halo, even though we can not directly observe its light.

So far, the redshift of this dark matter halo has not been constrained rigorously through lens modelling \citep[although approximations have been made in the past, see, e.g.][]{2021MNRAS.507.1662M}. In this paper, we aim to constrain its redshift, test the power of compound lensing to improve these constraints, and discuss whether or not a free redshift can alleviate the observed tension with $\Lambda$CDM. Furthermore, while previous studies have not yet done so, we include the first-order multipole perturbations (the "lopsidedness") in our mass model. We infer our posteriors using the open source lens modelling code {\sc Herculens\footnote{\href{https://github.com/Herculens/herculens}{https://github.com/Herculens/herculens}}} \citep[][]{2022A&A...668A.155G} in combination with stochastic variational inference (SVI) and a Hamiltonian-Montecarlo-within-Gibbs sampler \citep[][]{coleman_krawczyk_2024_12167630} using {\sc Numpyro} \citep[][]{phan2019composable,bingham2019pyro}.

The remainder of this paper is structured as follows. In Section \ref{sec:Data}, we describe the data that we consider for our analyses. In Section \ref{sec:StrongGravitationalLensing}, we provide a brief summary on compound lensing and outline the parameteric and pixelated models we use in this paper.
In Section \ref{sec:Inference}, we give a description of the statistical methods that we apply in this work. In Section \ref{Sec:Results}, we present our results, including constraints on the redshift of the halo, and discuss them in detail. Finally, we present our conclusions in Section \ref{sec:Conclusion}. 

\section{Data}
\label{sec:Data}

Our analysis considers the same data as \citetalias[][]{2024MNRAS.528.7564B}, i.e. an HST observation of \jackpot using the ACS in the F814W (I--band) with an exposure time of $t_{\rm exp} = 2096$\,s. This lens is part of the
 the SLACS sample  of lenses \citep{2008ApJ...677.1046G}. Unlike \citetalias[][]{2024MNRAS.528.7564B}, we do not subtract the lens light first, but rather model it simultaneously with the light from the sources and the mass models of the deflectors. The image has been drizzled to $0.05\arcsec$ per pixel \citep[see][]{2014MNRAS.443..969C}. We model image pixels of the data shown in Figure \ref{fig:s1_only_max_like_model}, and manually create masks for the arcs of the sources allowing for no overlap. 
 The lens system consists of a lens galaxy (which we will refer to as the main deflector) at redshift $z_{\rm main}= 0.222$, and three sources behind it (which we will also refer to as S1, S2, and S3). These sources are located at  redshifts $z_{\rm s1}=0.609$ \citep[][]{2008ApJ...677.1046G}, $z_{\rm s2}=2.045$ \citep[see][]{2021MNRAS.505.2136S}, and $z_{\rm s3}\approx 6$  \citep[see][]{2020MNRAS.497.1654C}.

\section{Strong gravitational lens modelling}
\label{sec:StrongGravitationalLensing}

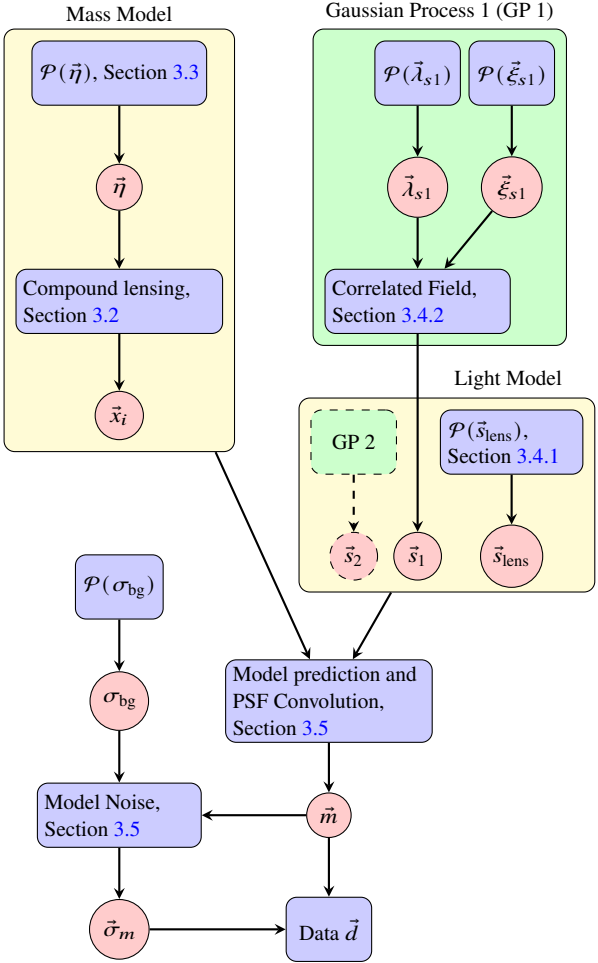
\begin{figure}
    \centering
    \begin{tikzpicture}[node distance=1.5cm]

\tikzstyle{block} = [rectangle, draw, fill=blue!20, text centered, rounded corners, minimum height=3em, minimum width=4em]
\tikzstyle{circleStyle} = [circle, draw, fill=red!20, text centered]
\tikzstyle{arrow} = [thick,->,>=stealth]
\tikzstyle{dashedblock} = [rectangle, rounded corners, draw, 
fill=yellow!20, inner sep=0.15cm, minimum height=5em, label={[label distance=2mm, above]:\textcolor{black}{}}]
\tikzstyle{dashedblock2} = [rectangle, rounded corners, draw, 
fill=green!20, inner sep=0.15cm, minimum height=5em, label={[label distance=2mm, above]:\textcolor{black}{}}]

\node [block] (input) {$\mathcal{P}(\vec \eta)$, Section \ref{sec:MassModel}};
\node [circleStyle, below of=input] (circle1) {$\vec \eta$};
\node [block, below of=circle1] (process1) {\parbox{9em}{Compound lensing, \\ Section \ref{sec:compound_lensing}}};
\node [circleStyle, below of=process1] (source_pos) {$\vec x_i$};
\node [block, right=2.2cm of input] (Plambda1) {$\mathcal{P}(\vec \lambda_{s1})$};
\node [circleStyle, below of=Plambda1] (lambda1) {$\vec \lambda_{s1}$};
\node [block, right=0.1cm of Plambda1] (white_noise1) {$\mathcal{P}(\vec \xi_{s1})$};
\node [circleStyle, below of=white_noise1] (xi1) {$\vec \xi_{s1}$};
\node [block, below of= lambda1] (source1) {\parbox{8em}{Correlated Field,\\ Section \ref{sec:pixelated_models}}};
\node [circleStyle, below=2.6cm of source1] (s1) {$\vec s_1$};  
\node [circleStyle, left=0.2cm of s1, dashed] (s2) {$\vec s_2$};  
 \node [block, fill=green!20, above of=s2, dashed] (gp2) {GP 2};    
\node [circleStyle, right=0.5cm of s1] (lens_light) {$\vec s_{\rm lens}$};
\node [block, above of=lens_light] (Plens_light) {\parbox{6em}{$\mathcal{P}(\vec s_{\rm lens})$,\\ Section \ref{sec:ParametricLightModel} } };
\node [block, below=1.5cm of source_pos] (Psigma_bg) {$\mathcal{P}(\sigma_{\rm bg})$};
\node [circleStyle, below of=Psigma_bg] (sigma_bg) {$\sigma_{\rm bg}$};
\node [block, below of=sigma_bg] (model_noise) {\parbox{7em}{Model Noise, \\ Section \ref{sec:model_observation}}};
\node [block, right=1cm of sigma_bg] (lensed_images) {\parbox{9em}{Model prediction and\\ PSF Convolution, \\ Section \ref{sec:model_observation}}};
\node [circleStyle, below of=lensed_images] (model) {$\vec m$};
\node [circleStyle, below of=model_noise] (sigma_model) {$\vec \sigma_m$};  
\node [block, below of=model] (data) {Data $\vec d$};

\begin{scope}[on background layer]
    \node[dashedblock, fit=(input) (source_pos) (circle1) (process1), label={[label distance=0mm, above]:\textcolor{black}{Mass Model}}] (box1) {};
     \node[dashedblock2, fit= (Plambda1) (source1)  (lambda1) (white_noise1) , label={[label distance=0mm, above]:\textcolor{black}{Gaussian Process 1  (GP 1)}} ] (gp) {};
      \node[dashedblock, label={[label distance=0mm, above right]:\textcolor{black}{Light Model}}, fit= (s1) (gp2) (s2) (Plens_light)  ] (box2) {};
\end{scope}

\draw [arrow] (input) -- node[anchor=east] {} (circle1);
\draw [arrow] (circle1) -- node[anchor=east] {} (process1);
\draw [arrow] (Plambda1) -- node[anchor=east] {} (lambda1);
\draw [arrow] (white_noise1) -- node[anchor=east] {} (xi1);
\draw [arrow] (process1) -- node[anchor=north] {} (source_pos);
\draw [arrow] (xi1) -- node[anchor=north] {} (source1);
\draw [arrow] (lambda1) -- node[anchor=east] {} (source1);
\draw [arrow] (box1) -- node[anchor=north] {} (lensed_images);
\draw [arrow] (source1) -- node[anchor=east] {} (s1);
\draw [arrow] (box2) -- node[anchor=east] {} (lensed_images);
\draw [arrow] (lensed_images) -- node[anchor=east] {} (model);
\draw [arrow] (Plens_light) -- node[anchor=east] {} (lens_light);
\draw [arrow] (model) -- node[anchor=east] {} (model_noise);
\draw [arrow] (model) -- node[anchor=east] {} (data);
\draw [arrow] (model_noise) -- node[anchor=east] {} (sigma_model);
\draw [arrow] (Psigma_bg) -- node[anchor=east] {} (sigma_bg);
\draw [arrow] (sigma_bg) -- node[anchor=east] {} (model_noise);
\draw [arrow] (sigma_model) -- node[anchor=east] {} (data);
\draw [arrow,dashed] (gp2) -- node[anchor=east] {} (s2);

\end{tikzpicture}
    \caption{The model components that we consider throughout this work. We further provide the sections, in which  these components are discussed in more detail. For simplicity we only show the Gaussian process (GP) that is used to model the first source in more detail. We use dashed lines to highlight that the second source is not always included.}
    \label{fig:model_overview}
\end{figure}

Below, we give a short summary of the main model components that we use throughout this work. We use {\sc herculens} and extend it for our purposes to compound lens systems. We further implement some additional lens mass profiles (see  e.g. Sections \ref{sec:multipoles} and \ref{sec:tNFW}). A diagram of our full model that we explain in the sections below is shown in Figure \ref{fig:model_overview}.

\subsection{Single plane lensing}

In this section, we provide a brief summary of the equations that describe strong gravitational lensing.
In gravitational lensing, the light of a background source (here a galaxy) is deflected by some intervening mass (here other galaxies). As a result, the image plane position, $\vec x$, at which we observe the light of a source, is displaced relative to its original position in projection, $\vec x_{\rm source}(\vec x)$, by some displacement $\DeflectionAngle$:
\begin{equation}
\vec x_{\rm source}(\vec x) = \vec x - \DeflectionAngle \,.
\end{equation}
The lensed images, $l(\vec x)$, depend on the both this displacement and the surface brightness of the source, $s(\vec x)$:
\begin{equation}
l(\vec x) = s(\vec x_{\rm source}(\vec x)) = s( \vec x - \DeflectionAngle  ) \,.
\end{equation}
Depending on the strength of the displacement, the resulting images can be highly distorted and magnified. Strong lensing occurs when the displacement is strong enough to create multiple images of the background source. 

The displacement $\alpha(\vec x)$ is related to the lensing potential, $\Psi(\vec x)$,  and convergence, $\kappa(\vec x)$, according to:
\begin{equation}
\Sigma(\vec x)/\Sigma_{\rm crit}  = \kappa(\vec x)  = \nabla  \DeflectionAngle  /2 = \nabla^2 \Psi(\vec x) /2\,,
\label{equ:disp_conv_psi}
\end{equation}
with $\Sigma(\vec x)$ being the projected mass density of the lens and $\Sigma_{\rm crit}$ being the critical surface density. $\Sigma_{\rm crit}= \frac{c^2}{4\pi G}\frac{D_{s}}{D_{d,s}D_{d}}$ depends on the angular diameter distances between and towards the deflector and source, that is $D_{\rm d}$, $D_{\rm s}$, and $D_{d,s}$.  As these distances depend on cosmological parameters, we will assume $\Omega_m=1-\Omega_\Lambda=0.3103$ 
 and $h=0.6766$ \citep[taken from][]{2020A&A...641A...6P} in a flat $\Lambda$CDM cosmology.  
Calculating the displacement for a given mass distribution is equivalent to solving Equation \ref{equ:disp_conv_psi} for $\alpha(\vec x)$, usually for a given choice of the convergence, $\kappa(\vec x)$.

\subsection{Compound lensing}
\label{sec:compound_lensing}
The lens system we study in this paper does not only have a single lens, but multiple galaxies that act as lenses on the sources behind them. In this compound lensing scenario, the lens equation is recursive and the projected position on the $j$-th redshift plane, $\vec x_j$, becomes:
\begin{equation}
    {\vec x}_{j}=\vec x_{0}-\sum_{i=1}^{j} \hat \beta_{ij} \times {\alpha}_{i-1}({\vec x}_{i-1}) \text{ for j > 0} \,.
\label{eq:multiplanelensequation}
\end{equation}
 The parameters $\hat  \beta_{ij}$ are related to the family ratios \citep[see e.g.][]{2018ApJ...860...94G}, and are determined from ratios of angular diameter distances between the different redshift planes:

\begin{equation}
\hat  \beta_{ij} = \frac{D_{i,j} D_{i+1} }{D_{i} D_{i,i+1}}\,.
\end{equation}
We choose $\hat  \beta_{ij}$ such that convergences are defined relative to the next plane in redshift. We denote the sum over the images created by all the redshift planes as  $l(\vec x_0) = \sum_{j=0}^{N_{z}} s_j(\vec x_j)$. 

\subsection{Mass model}

\label{sec:MassModel}

\subsubsection{Elliptical Power law and external Shear}
     We assume that to first order lenses are well described by an elliptical power law (EPL) mass profile, which we parameterize as:
    \begin{equation}
        \kappa_{\rm EPL}(\rho,\gamma,q) = \frac{3-\gamma}{2} \left(\frac{\theta_E}{\rho}\right)^{\gamma-1} \,,
    \end{equation}
    where $\rho^2=x^2q+y^2/q$ is the squared ellitpical radius. We further allow the center $x_c,y_c$ and position angle, $\theta$, to vary freely.
    Rather than choosing a prior on the minor-to-major axis ratio $q$ and position angle $\theta$, we choose a prior on the ellipticities $\vec e = (e_x,e_y) = \frac{1 - q}{1+q}\times ( \cos(2\theta), \sin(2\theta) )$ as described in {\sc Herculens} \citep[][]{2022A&A...668A.155G} and {\sc Lenstronomy} \citep[][]{,2018PDU....22..189B}. 
    In addition to this power-law mass model we allow for an external shear component, characterized by $\Gamma_x$ and $\Gamma_y$. 
    We denote the collection of EPL and Shear parameter as $\vec \eta_{\rm EPL}$ and $\vec \eta_{\rm \Gamma}$.

 We always model the main deflector at $z_{\rm main}=0.222$ with an EPL with external Shear. Whenever Source 2 is included, we further model the first sources mass distribution as an singular isothermal ellipse ($\gamma = 2$), for which the center is fixed to where the average of four conjugate points is traced to on the first source plane   \citepalias[following][]{2024MNRAS.528.7564B}.

\subsubsection{Multipoles}
\label{sec:multipoles}

To account for additional complexity in the lens, we include multipole perturbations in our model of the main deflector. We parameterize them in a way that includes the slope $\gamma$ of the power-law profile and consider the orders $n \in \{1,3,4\}$ \cite[implemented similar to][ such that the slope is matched to the main deflector]{2022MNRAS.516.1808P}. The convergence that is associated with a given multipole is:
\begin{equation}
     \kappa_{\mathcal{M}_n}(\rho,\gamma,a_n,\phi_n) =  \kappa_{\rm EPL}(\rho,\gamma, 1) \times  A_{\mathcal{M}_n} \times \cos(n(\phi - \phi_n)) \,,
\end{equation}
where $ A_{\mathcal{M}_n} $ is the amplitude of the $n$-th order multipole perturbation and $\phi_n$ determines its orientation \cite[see e.g.][]{2013ApJ...765..134C}.
Similar to the axis ratio $q$ and position angle $\theta$ 
of the EPL, we do not construct our prior in terms of  amplitude $A_{\mathcal{M}_n}$ and angle $\phi_n$. Instead we define the multipole equivalent of ellipticities $\vec e_n$, such that $A_{\mathcal{M}_n} = 1 - q_n$ and $\phi_n =  \theta_n \times 2 / n $. In particular, we choose $\vec e_n = \frac{1 - q_n}{1+q_n}\times ( \cos(2\theta_n), \sin(2\theta_n) )$. This ensures that the periodic boundary conditions are fullfilled, but also provides reasonable priors on the amplitudes of the multipoles.

Recent work has shown the importance of multipole perturbations to correctly recover subhaloes  in lens modelling \citep[see e.g. ][]{2024MNRAS.52710480N, 2024MNRAS.528.1757O}, as the additional complexity can be degenerate with the presence of dark matter haloes \citep[in particular the $\mathcal{M}_1$ multipole, see e.g.][]{2024arXiv240712983A,2024arXiv241012987L}. Multipole perturbations have further been shown to affect the analysis of flux ratios from quasars \citep[see e.g.][]{2024MNRAS.531.3431C}.

\subsubsection{Dark halo perturber}
\label{sec:tNFW}

We assume that the perturber has a truncated NFW density profile  \citep[][]{1997ApJ...490..493N}. The three dimensional NFW profile is:
\begin{equation}
\rho^{\rm 3D}_{\rm NFW}(r) =   \frac{\rho_s  }{ 
\frac{r}{r_s} (1+ \frac{r}{r_s})^2 }\,.
\end{equation}
In projection this gives rise to a convergence of the following form \cite[see e.g.][]{2001astro.ph..2341K}:
\begin{equation}
\kappa_{\rm NFW}(r) =  2 \kappa_s \frac{ 1.0 - F(\frac{r}{r_s}) }{ \left(\frac{r}{r_s}\right)^2  - 1 }\,,
\end{equation}
with $r^2= x^2+y^2$ , $r_s$ being the scale radius of the NFW profile, and $k_s$ determining the amplitude of the convergence.\footnote{ The expression above uses $ F(x) = \tan^{-1} \left(\sqrt{x^2-1}\right) / \sqrt{x^2-1}$ for $x>1$ and $\tanh^{-1} \left(\sqrt{1-x^2}\right) / \sqrt{1-x^2}$ for $0<x\leq1$.}

We further allow the halo to be truncated at a radius $r_t$. We include this additional parameter to capture the effects of tidal stripping that a subhalo could experiences when interacting with its host. As pointed out by \citetalias[][]{2024arXiv240811090M}, it can further capture the virial radius at in-fall, as it would be expected to be smaller for a subhalo than for a field halo. This profile further allows easy comparison with previous works. We follow \citet[][]{2009JCAP...01..015B} and \citet[][]{2011MNRAS.414.1851O} and assume a truncated NFW profile of the following form:
\begin{equation}
    \rho_{\rm tNFW}^{\rm 3D}(r)  = \rho_{\rm NFW}^{\rm 3D}(r) \times \frac{r_t^2}{r_t^2+r^2} \,.
\end{equation}

The combination of all parameters that describe the perturber is referred to as $\eta_{\rm tNFW}$.  The halos center coordinates are defined relative to $x_0=-0.68$, $y_0=1.0$ (or where this position is mapped to on the lens plane, if the subhalo is behind the main deflector).  This position falls close to where the dark halo has been found in previous studies \citepalias[see e.g.][]{2024MNRAS.528.7564B}.

\begin{figure*}
    \centering
    \includegraphics[width=0.6\textwidth]{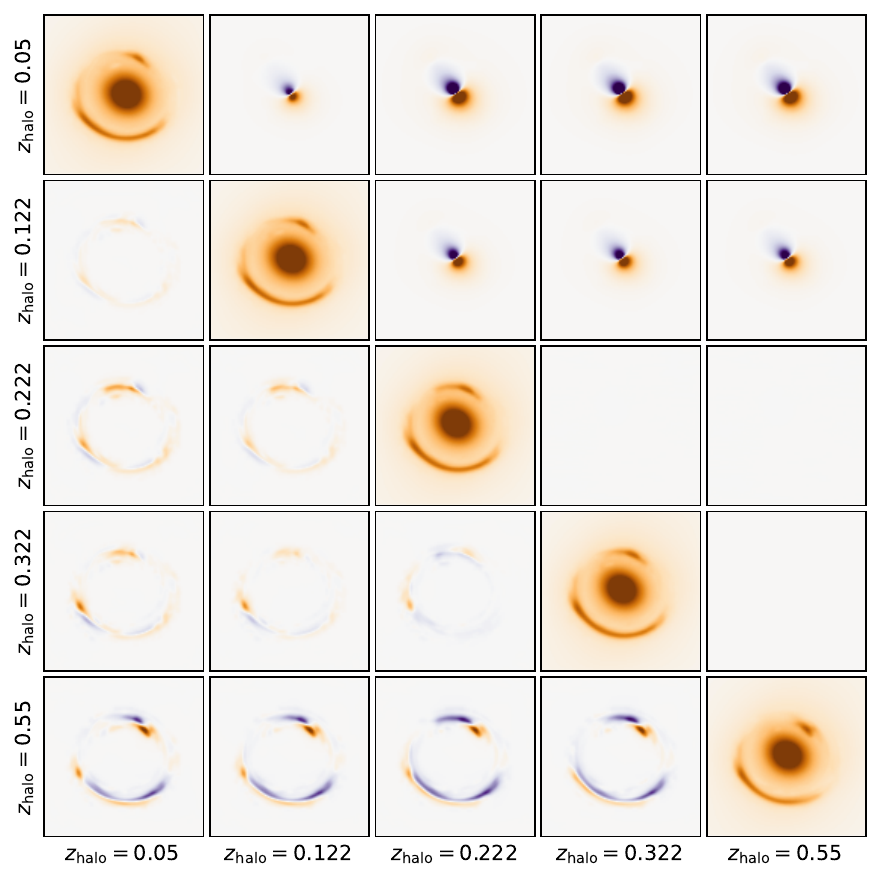}
    \caption{The effects that the subhalo redshift has on the lensed images of a background source and the main deflector light. We show five mock observations on the diagonal with halo redshifts $z_{\rm halo} \in \{0.05,0.202,0.222,0.322,0.55\}$. All other parameters are fixed to the same values. The upper right triangle shows the differences between each combination of those mocks, considering only the light of the main deflector. The lower left triangle shows the differences between each combination of those models, considering only the resulting arcs that they generate.}
    \label{fig:redshift_differences}
\end{figure*}

In this paper, we are interested in inferring the redshift of this dark matter halo,  $z_{\rm halo}$. To visualize the effects that a free redshift mass component can generate, Figure \ref{fig:redshift_differences} shows the observed light distribution of mock data for different halo redshifts in front, behind, and at the main deflector. All other parameters are kept fixed to emphasize these effects. The massive and concentrated perturber creates additional magnification and distortion in the observed arcs for all redshifts $z<z_{s1}$. Still, it will only affect the lens light if placed sufficiently far in front of it, where its lensing effect becomes efficient. We note that in practice, differences will be much smaller after lens modelling because changes in the main deflector or source light distribution can absorb some of the illustrated differences. Even so, compound lensing can generate lensed images that are  qualitatively very different to those generated by single-plane lenses \cite[see e.g.][]{2016MNRAS.456.2210C}.

\subsection{Light model}
\subsubsection{Parametric lens light}

\label{sec:ParametricLightModel}

We assume that the lens light can be constructed from a sum of Gaussian components \cite[see e.g.][]{2019MNRAS.488.1387S,2024MNRAS.532.2441H}, each parametrized as:
\begin{equation}
\mathcal{G}(\rho_n, \sigma_n) = \frac{A_n}{2\pi } \exp \Big( - \frac{1}{2} \rho_n^2 / \sigma_n^2 \Big) \,, 
\end{equation}
where we allow the amplitudes $A_n$, the axis ratios $q_n$ that appear in the radius $\rho_n=x^2q_n+y^2/q_n$,  centers and position angles to be free parameters for each component individually. 

Swapping the parameters of two Gaussian components can in principle provide the same light model. To avoid additional modes in the posterior arising from this symmetry, we restrict the different Gaussian components to mutually exclusive ranges for their width $\sigma_n$. The indivdual ranges of the $N_{\rm gauss} = 20$ components are uniformly spaced in the logarithm of $\sigma$ \citep[similar to][]{2002MNRAS.333..400C,2019MNRAS.488.1387S}, i.e. the range of the $n$-th component is defined such that $\log( \sigma_n / \sigma_{\rm min}) / \Delta \sigma \in [ n , n+1]$ with $\Delta \sigma = \frac{\log(\sigma_{\rm min}/\sigma_{\rm max})}{N_{\rm gauss}-1}$. We refer to the combination of all those parameters as $\vec s_{\rm lens}$. We choose $\sigma_{\rm min} = 0.01 $ and $\sigma_{\rm max} = 3.00$, which provides a good fit to the image.

\subsubsection{Pixelated Sources}
\label{sec:pixelated_models}
We assume the background sources $s(\vec x)$ are fields that vary on regularly pixelated grids. The grid on which we reconstruct sources is fixed to 
the smallest square containing all the source positions corresponding to the image pixels within the masks shown in Figure~\ref{fig:s1_only_max_like_model}.

We assume that these fields are reasonably well described by a Gaussian process (GP). The field values in each pixel are the components of the vectors $\vec s$. For a GP, these values follow Gaussian statistics, such that in Fourier space  the amplitudes are related to a power spectrum \citep[see e.g.][]{2022A&A...668A.155G,2024A&A...682A.146R}.
\begin{equation}
\vec { s} = \mathcal{F} \sqrt{P} \odot \vec \xi \,,
\end{equation}
with each element of $\vec \xi$ being drawn from a zero mean unit variance Normal distribution, $\odot$ being the pointwise multiplication, and $\mathcal{F}$ being the Fourier transform. Following the regularization approach by \citep[][]{2006MNRAS.371..983S}, the previous analysis of \jackpot by \citetalias[][]{2024MNRAS.528.7564B} has shown that gradient regularization is preferred over curvature regularization.
In this work, we want to the data to determine which regularization is used. To do so we choose a Mat\'ern power spectrum \citep[see e.g.][]{stein2012interpolation} similar to previous works \citep[such as][]{2020MNRAS.499.5641V,2022MNRAS.516.1347V,2024arXiv240608484G,2024A&A...682A.146R}:

\begin{equation}
    P_s(k) = \sigma^2 \times  4 \pi n \left(\frac{2n}{\zeta^2}\right)^n\times  \left(\frac{2n}{\zeta^2} + k^2 \right)^{-(n+1)}\,.
\end{equation}

The Matern power spectrum is well-suited for a large variety of random fields. For small values of of $\frac{2n}{\zeta^2}$ this power spectrum becomes similar to the more classical regularization schemes. A power spectrum of the form $P(k) = k^{-2n}$ corresponds to regularization of the $n$-the derivative, i.e. $n=1$ corresponds to gradient and $n=2$ to curvature regularization. We note that we define our power spectrum parameters in pixel units to avoid the artificial breaking of the mass sheet degeneracy. The combination of all parameters that define a source power spectrum are denoted by $\vec \lambda = (n,\zeta, \sigma)$.

\subsection{Model observations}
\label{sec:model_observation}

\begin{figure*}
    \centering
    \includegraphics[width=\textwidth]{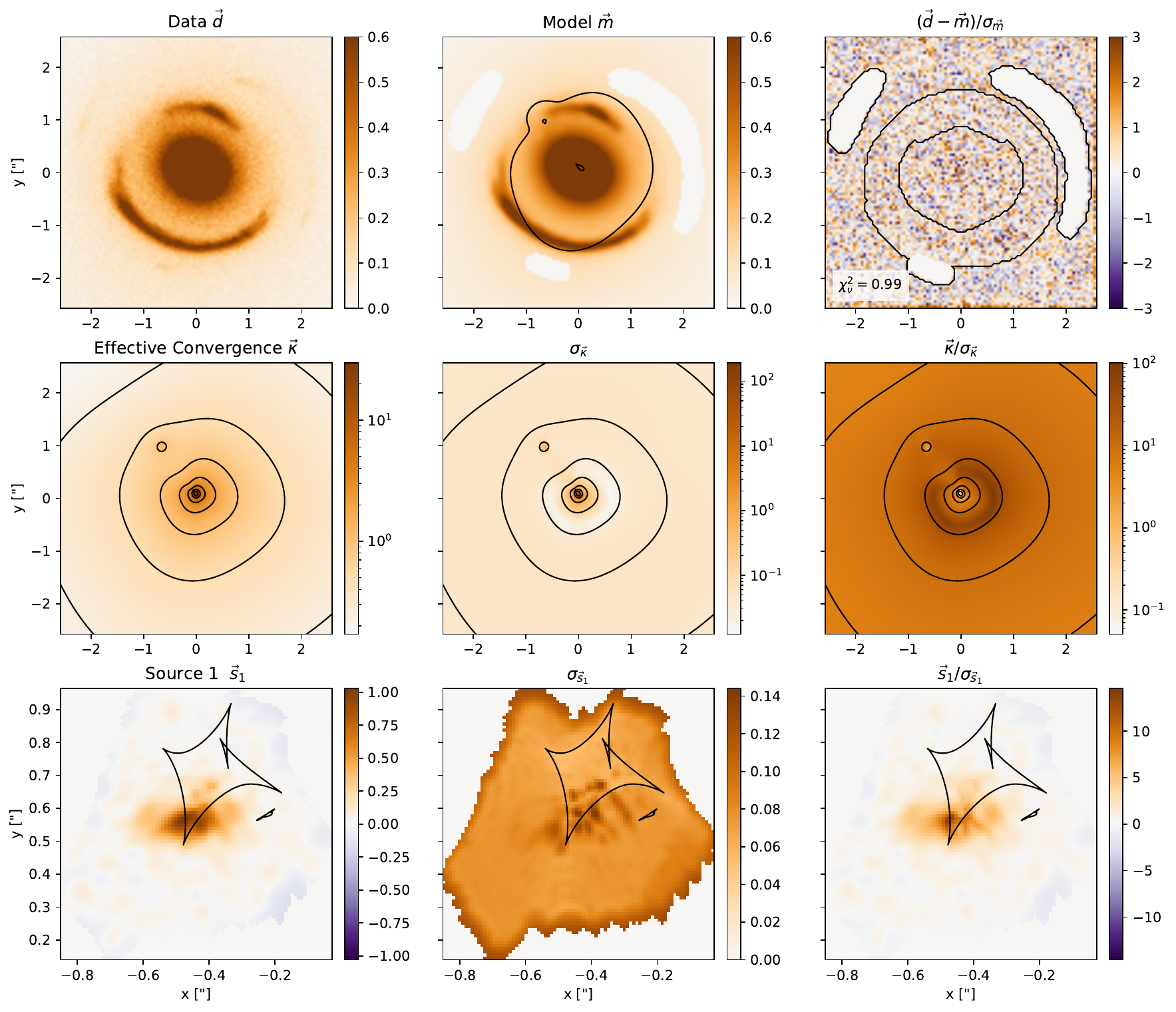}
    \caption{The mean posterior S1-only model. Shown are the original data, our model predictions, the noise-weighted residuals (row 1), the effective convergence (row 2), and the reconstructed Source 1 (row 3). The figures reporting the ratio of mean and standard deviations highlight the features most constrained by the data given our model assumptions. We also report the pixel averaged residual with $\chi^2_{\nu}$. Note that we show the effective convergence, since the redshift of the dark halo is allowed to change  during our reconstruction \citep[see e.g.][]{2020MNRAS.491.6077G}.}
    \label{fig:s1_only_max_like_model}
\end{figure*}

\begin{figure*}
    \centering
    \includegraphics[width=\textwidth]{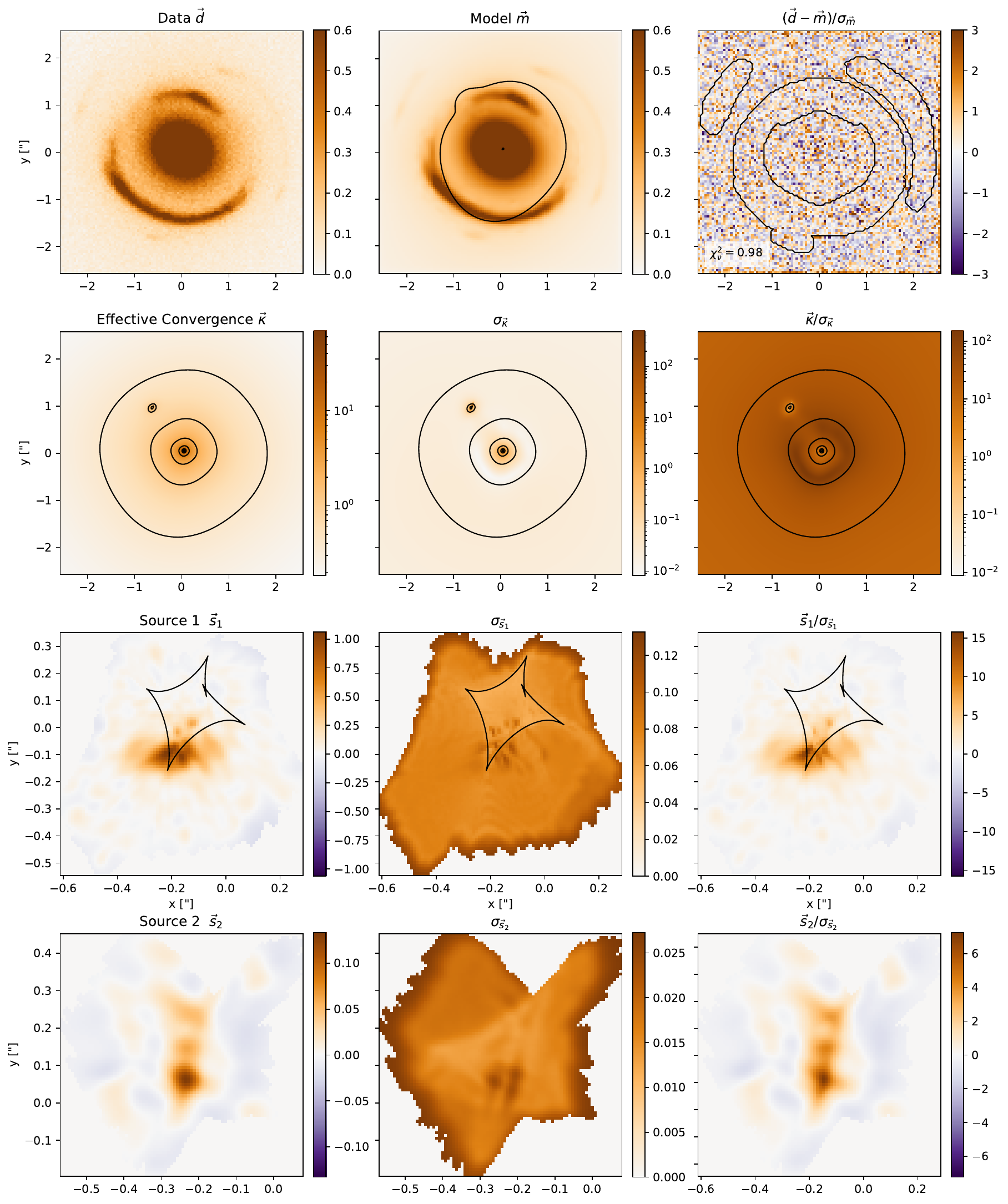}
    \caption{Same as Figure \ref{fig:s1_only_max_like_model} but for our S1\&S2 model. The second source is shown in the fourth row. }
\label{fig:s1_s2_max_like_model}
\end{figure*}

Our analysis includes a point-spread-function (PSF) determined from another star that lies nearby the lens in consideration \citep[taken from][]{2014MNRAS.443..969C}. The noiseless model is then given by the sum of all light components (lens and lensed source images) convolved with the PSF, i.e. $m(\vec x) = \int d\vec y b(\vec x- \vec y) l(\vec y)$. To evaluate the model $m(\vec x)$ we use {\sc Herculens}  \citep[][]{2022A&A...668A.155G}, that we extended to compound lensing (and the mass profiles as described above if they were not yet implemented).
We explicitly set the source brightness to zero outside of the masks (shown in Figures \ref{fig:s1_only_max_like_model} and  \ref{fig:s1_s2_max_like_model}) before the lensed images are being convolved. This assumption is justified, since we do not expect the sources to contribute to the observed light distribution anywhere outside of these masks.
We evaluate the convolution by multiplication of the Fast Fourier Transforms of the PSF and the lensed images, while using zero-padding to avoid artifacts from periodic boundary conditions.

We include in our model the fact that different realizations of the model will give rise to different noise properties. In particular, we assume zero mean gaussian noise, $\vec n$, with a standard deviation of:
\begin{equation}
    \vec \sigma_m^2 = \sigma_{\rm bg}^2 \vec 1 + \vec m(\vec \eta, \vec s, z_{\rm halo}) / t_{\rm exp} \,,
\end{equation}
thereby accounting for the fact that the noise will depend on the signal in each pixel.
Here, we defined the vectors $\vec s = (\vec s_{\rm lens}, \vec s_1, \vec s_2, \vec \lambda_{s1}, 
\vec \lambda_{s2} )$ and $\vec \eta=(\eta_{\rm EPL,lens},\eta_{\mathcal{M}_1},\eta_{\mathcal{M}_3},\eta_{\mathcal{M}_4},\eta_{\rm tNFW},\eta_{\rm EPL,s1} )$.  We infer the background noise level $\sigma_{\rm bg}$ assuming that it is drawn from a half normal distribution with a variance matching the one of background pixels in the corners of our cutout.
While our data has been drizzled, we do not account for the spatially correlated noise that would result from that.
For each data pixel we use a supersampling factor of four as implemented in the base version of {\sc Herculens}. Recently, \citetalias[][]{2024arXiv240811090M} has shown that this is particularly important for the detection of perturbers. We create grids with $86\times86$ pixels for both sources, as this provides a sufficient coverage of raytraced points per pixel. \footnote{ \nopagebreak In this context, sufficient coverage refers to the Nyquist frequency, which corresponds to the highest frequency that can be accurately recovered from a sampled signal.  To achieve this we adjust grid sizes until on average two rays from the supersampled image plane trace back to each source pixel of interest.}

\begin{figure*}
    \centering
    \includegraphics[width=0.7\textwidth]{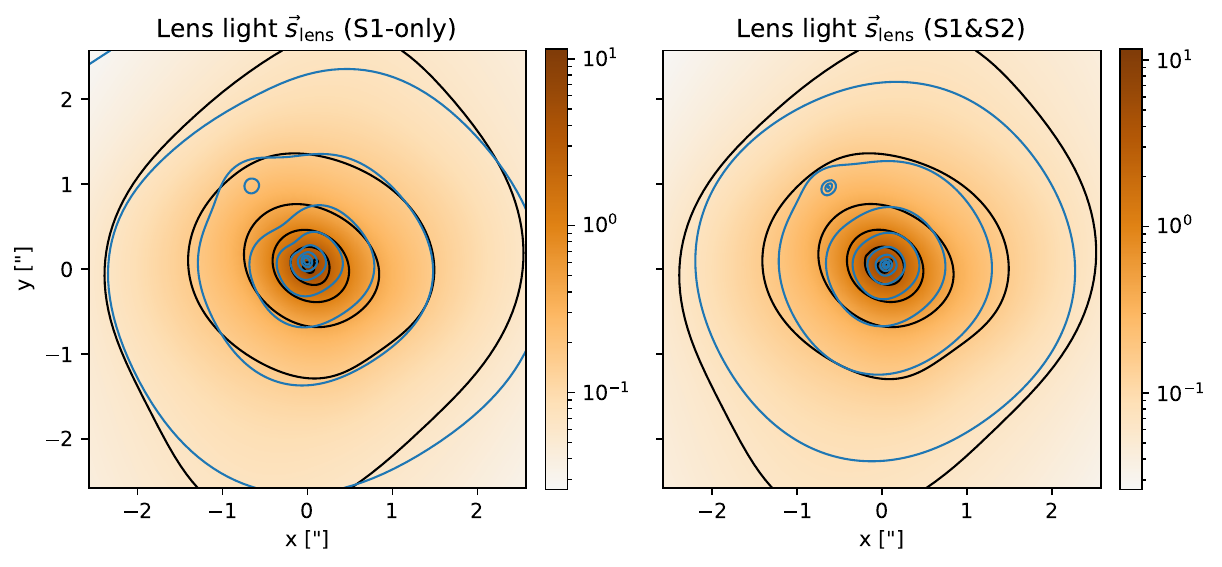}
    \caption{The reconstructed light distribution of the main deflector for our S1-only model (left) and our S1\&S2 model (right). In each panel, we show the isophotes  of the main deflector (black) to highlight the change in the orientation of the major axis with radius. We further include contours of the reconstructed mass distribution (blue) for comparison.}
    \label{fig:lens_light}
\end{figure*}

\begin{figure*}
    \centering
    \begin{overpic}[width=\textwidth]{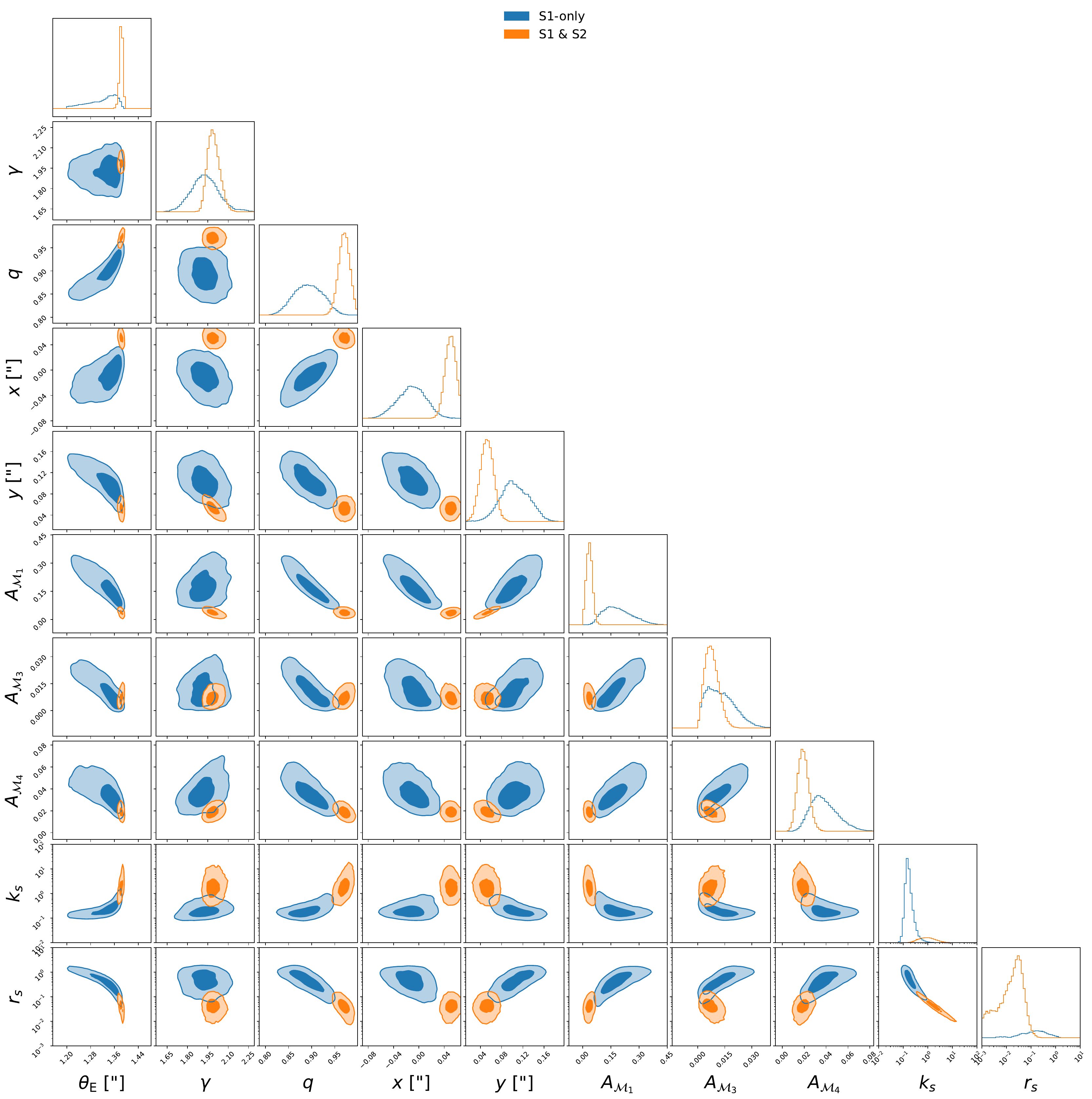}
    \put(53.5, 53.5){\includegraphics[width=0.46\textwidth]{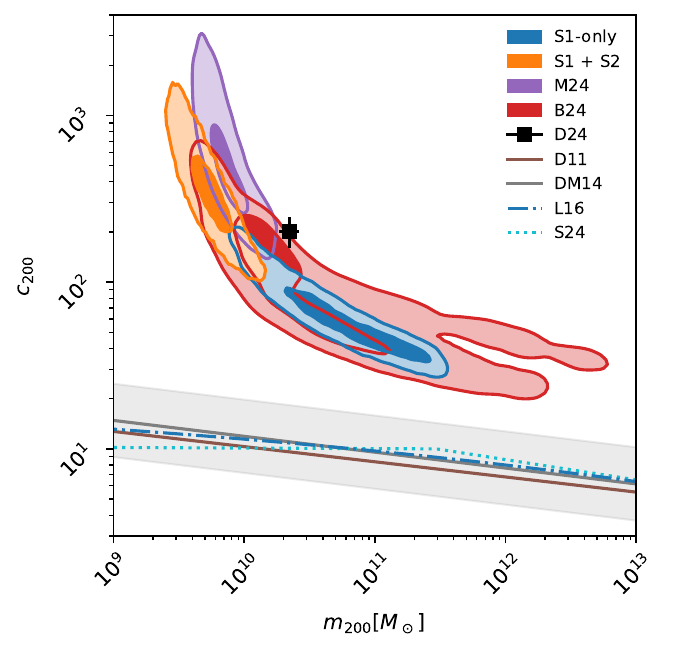}} 
    \end{overpic}
    \caption{The 1 and $2\sigma$ posterior contours of a subset of model parameters. Blue shows our single plane model, whilst orange is the two source plane fit. The results emphasize how strongly the second source breaks model degeneracies.
    The top right panel further shows our posterior constraints on the mass and concentration of the dark halo, and  compares to previous studies \citetalias[][]{2024MNRAS.528.7564B}, \citetalias[][]{2024arXiv240712910D},  and \citetalias[][]{2024arXiv240811090M}. 
  We further show the $\Lambda$CDM expectations from \citet[][D11]{2008MNRAS.390L..64D} ,  \citet[][DM14]{2014MNRAS.441.3359D} , \citet[][L16]{2016MNRAS.460.1214L}, and \citet[][S24]{2024arXiv240901758S}.    }
    \label{fig:triangle_main_cm-relation}
\end{figure*}

\section{Bayesian Inference}
\label{sec:Inference}

In the previous sections, we described  all the parameters that enter our models.
Below, we will first discuss the posterior, from which we aim to sample from, followed by a brief discussion of the sampling strategy we employ for this purpose.

\subsection{Lens modelling posterior }

The posterior from which we want to draw samples:
\begin{equation}
\mathcal{P}(\vec \eta, \vec s, \sigma_{\rm bg}^2 |\vec d) = \frac{ \mathcal{P}( \vec d |\vec \eta, \vec s, \sigma_{\rm bg} ) \mathcal{P}(\vec \eta, \vec s, \sigma_{\rm bg}, z_{\rm halo}) } { \mathcal{P}(\vec d)  } \,.
\end{equation}
Under the assumption of Gaussian noise the Likelihood becomes:
\begin{equation}
    \mathcal{P}( \vec d | \vec \eta, \vec s, \sigma_{\rm bg}^2)  = \mathcal{G}  \left( \vec d - \vec m(\vec \eta, \vec s, z_{\rm halo}) , \textrm{diag} \left(\vec \sigma_m^2 \right) \right) \,,
\end{equation}
where the $\textrm{diag}$ operator creates a matrix with the variance in each pixel along the diagonal. In Tables \ref{tab:HMCgibbsMain} and \ref{tab:HMCgibbsSub}, we report our choices of priors for the target parameters or refer to the relevant sections of this paper. While the Bayesian evidence is interesting for model comparison, we will not explicitly calculate it within this work and leave this up to future work.

We aim to obtain samples from the posterior described above. This is a challenging task because of several reasons: 

\begin{enumerate}

    \item 
    First of all, it is an inverse problem in the sense that we can easily model the generation of data in a forward fashion. Still, no explicit inverse function exists that returns the parameters originally entering the model that gave rise to a realization of data. Simultaneously reconstructing the source light and lens mass is also an ill-posed problem, i.e. multiple combinations thereof can give rise to the same observed data. One example of this is the well-studied Mass-Sheet Transformation (MST) or more general Source Position Transformations \citep[][]{2014A&A...564A.103S}. 
This highlights the importance of simultaneously fitting for the light and mass models to avoid artificially breaking some of these degeneracies.

    \item Lens models (mainly w.r.t. the mass model, but in the case of signal dependent noise also w.r.t. the sources) are often highly non-linear. This becomes even more challenging in the case of compound lensing as shown in Equation (\ref{eq:multiplanelensequation}).  Furthermore, lens models show an intrinsic hierarchical structure, in which the source light distribution is evaluated depending on the mass distribution of the deflector. This is challenging, since hierarchies tend to generate funnel-shaped likelihoods (with a large variance in one part and an increasingly smaller variance in another part of the parameter space) that  samplers can struggle with \citep[see e.g.][]{2000physics...9028N}. Methods such as Slice-sampling or Gibbs sampling can alleviate such challenges.

    \item Finally, the reconstruction of pixelated sources or lensing potentials renders lens modelling a high-dimensional problem. Accordingly, one requires sampling strategies that will not suffer from low acceptance probabilities \citep[e.g. Hamiltonian Monte Carlo methods, HMC,][]{2011hmcm.book..113N}.
\end{enumerate}

\subsection{Sampling strategy}

To address these challenges, we employ auto-differentiable programming using  the {\sc Jax}-based \citep[][]{jax2018github} software {\sc Herculens} \citep[][]{2022A&A...668A.155G} in combination with the probabilistic programming package {\sc Numpyro} \citep[][]{phan2019composable,bingham2019pyro}. Our analysis follows a forward modelling philosophy, i.e. for a given set of target and nuisance parameters our model can generate simulated data according to the process shown in Figure \ref{fig:model_overview}. This simulated data can then be compared to real data to draw conclusions about the input parameters.
Our analysis happens in two stages:

\begin{enumerate}
    \item We first make use  of {\sc Numpyro}'s implementation of Stochastic Variational Inference \citep[SVI, see e.g.][]{2013arXiv1301.1299W} to find an approximation to the target posterior distributions. The SVI is computationally much cheaper than Markov Chain Monte Carlo methods \citep[see e.g.][]{2021Entrp..23..853F,2024arXiv240608484G} as it can make use of the auto-differentiable nature of {\sc Jax} to minimize its loss function. We employ the {\sc AdaBelief} optimizer \citep[][]{2020arXiv201007468Z} for this purpose. 
    We choose a low-rank multivariate Normal distribution as the guiding probability distribution, which allows to capture the main features we expect to appear in the true posterior.\footnote{We note that  {\sc Numpyro} 
    performs the SVI in unconstrained spaces of the target parameters and therefore includes some coordinate transformations. This  renders it possible to capture at least some non-gaussian features in the resulting SVI approximation.}
    We incrementally allow for more complexity of the model in this step (employing parameteric sources before we move on to pixelated sources). The loss function of the SVI is the Evidence Lower Bound (ELBO), which can be used to explore different models early on in the analysis with an indication for their suitability. 
    While the SVI is ideal for exploring different model assumptions, it tends to underestimate uncertainties, is mode seeking rather than mode covering (at least for the implementation of ELBO considered here), and is limited by the assumed shape of the approximate posterior.

\item To obtain more robust results, we employ the HMC-within-Gibbs sampler of \citet[][]{coleman_krawczyk_2024_12167630} together with the No U-Turn Sampler \citep[NUTS,][]{2011arXiv1111.4246H} implemented in {\sc Numpyro}. This setup allows us to draw samples from our posterior in a Gibbs sampling fashion \citep[][]{gelman2013Bayesian}:
we sequentially update a subset of the overall model parameters while keeping the remaining parameters fixed. This approach renders our analysis much more robust in light of the hierarchical nature of lens modelling. Similar to  \citet[][]{2022ApJ...935...49G}, we precondition our HMC by choosing the inverse covariance matrix of the SVI posterior as the initial mass matrix. 
We subdivide the parameters of our model into several blocks for our single (double) source models. We split our model parameters into the several subsets: $\vec s_{\rm lens}$,
   ($\vec s_1$, $\vec \lambda_{s1}$), ($\vec s_2$, $\vec \lambda_{s2}$),
   ($\vec \eta_{\rm EPL,lens}$, $\vec \eta_{\mathcal{M}_1}$, $\vec \eta_{\mathcal{M}_3}$, $\vec \eta_{\mathcal{M}_4}$ $\vec \eta_{\rm EPL,s1}$, $\eta_{\rm tNFW}$, $\sigma_{\rm bg}$ ),
   $z_{\rm tNFW}$. 
Within a single Gibbs step we subsequently sample the parameters of one of these subsets while conditioning on the remaining parameters. Sampling once over each of the conditional posteriors generates one new sample of our target posterior. The above choice of subsets ensures that our  mass model is in each step sampled according to the light models.

\end{enumerate}

For both the SVI and HMC-within-Gibbs step we consider 4 independent chains with different initial random random seeds.  We generate 20k posterior samples per chain in the second stage, each drawn after 6000 steps of warm-up. We perform analyses for both a single (S1-only) and a double source plane (S1\&S2) model.

\section{Results and Discussion}
\label{Sec:Results}

In the following, we present the results of our analyses with the model outlined in Section \ref{sec:StrongGravitationalLensing}. Figures \ref{fig:s1_only_max_like_model}, \ref{fig:lens_light} and \ref{fig:s1_s2_max_like_model} show our mean reconstructed S1-only and S1\&S2 models. In both cases we reconstruct the data up to the noise level. In what follows, we will discuss these results in more detail.

\subsection{A main deflector with significant multipoles}

We find that most main deflector parameters are compatible with the recent studies  from \citetalias[][]{2024MNRAS.528.7564B} and \citetalias{2024arXiv240811090M}, matching their parameters within $2\sigma$. To highlight this, we show in Table \ref{tab:HMCgibbsMain} the number of standard deviations within which our posterior medians agree. Some larger differences appear in the external Shear and are probably explained by the inclusion of the $\mathcal{M}_1$ multipole, which shows the highest amplitude of all multipoles in our reconstruction with $A_{\mathcal{M}_1}={0.036}^{+0.016}_{-0.016}$ for our S1\&S2 model. Since the inclusion of external shear can alleviate some deficiencies in the mass model, rather than reflecting a truly external shear \citep[see][]{2024MNRAS.531.3684E}, we do not expect the difference in shear to present a problem in our analysis.

Overall we find that the mass distribution of \jackpot is more boxy with the corners roughly aligning with those of the light distribution shown in Figure \ref{fig:lens_light}. This points towards the mass distribution following the lens light more closely than a simple EPL model would allow for. In agreement with \citet[][]{2024MNRAS.532.2441H}, we find that the lens light is very well described by a sum of Gaussian components. While the uncertainty of the lens light is higher close to the center, we find that it is strongly constrained independent of the inclusion of the second source. We, therefore, expect that our lens light subtraction is robust, which is important for the identification of dark (sub)haloes \citep[see e.g.][]{2024MNRAS.52710480N}.
The orientation of Gaussian ellipses changes between the inner and outer parts of the lens light distributions. Together with the mass distributions this is characteristic for a galaxy that previously experienced a merger or interaction with another galaxy. Our constraints of $A_{\mathcal{M}_1} = 3.6$ percent are consistent with the expectation from such interaction, which is on the order of $2$ percent \citep[at least for the light distributions of low redshift galaxies, see e.g.][]{2024arXiv240712983A}. Our inferred $A_{\mathcal{M}_1}$ is also similar to the one found for another lens system that has recently been studied by \citet[][]{2024arXiv241012987L}, which shows $A_{\mathcal{M}_1} = 2.4-9.5$ percent.

Several degeneracies exist in the posterior excluding the second source. Figure \ref{fig:triangle_main_cm-relation} shows some of the most affected posterior parameters. Notable is the strong degeneracy between the first order multipole ($A_{\mathcal{M}_1}$) and the center of the main deflector. 
In agreement with \ballard, many of these degeneracies disappear with the inclusion of the second source in the model, as the arcs of the second source probe the mass distribution up to higher radii.
This is not unique to compound lensing, as multiple background sources at the same redshift would also lead to additional constraints.

\subsection{A very cuspy dark matter halo at the main deflector redshift}

\begin{figure}
    \centering
    \includegraphics[width=\columnwidth]{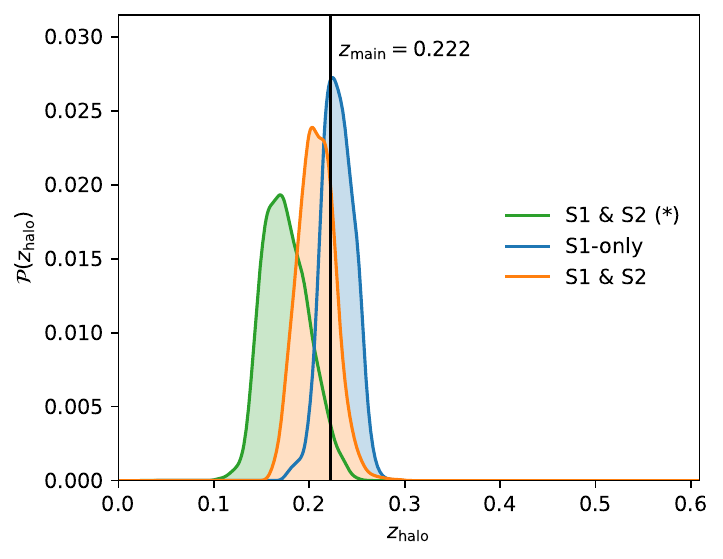}
    \caption{The posterior redshift distribution from our single source  (blue) and double source (orange) lens models. The green S1 \& S2 (*) shows a model where the light of the main deflector is not affected by the dark halo, even when it is in front of the lens.
    }
\label{fig:redshift_posterior}
\end{figure}

Figure \ref{fig:redshift_posterior} shows our halo redshift constraints, which for our S1-only model become $z_{\rm halo}={0.229}^{+0.018}_{-0.015}$. The inclusion of the second source plane leads to no significant difference on the redshift uncertainty, but rather a small shift, i.e. $z_{\rm halo}={0.207}^{+0.019}_{-0.019}$. Both posteriors point towards the halo being a subhalo. 
A ballpark estimate for the preference of the subhalo model over a field halo can be obtained from the ELBOs of models with and without a free redshift. Assuming each model is equally likely, the ratio of the ELBOs (corresponding to the Bayes factor $\alpha_{\rm Bayes}$) prefers a subhalo with $\log \alpha_{\rm Bayes} \approx 8$. However, \citet{2018MNRAS.475.5424D} showed that a line-of-sight halo was apriori twice as likely as a subhalo. Putting these together, we find $\log \alpha_{\rm Bayes} \approx 7$, corresponding to a 1 in 1000 chance of a field halo rather than a subhalo.

\begin{figure*}
    \centering
    \includegraphics[width=\textwidth]{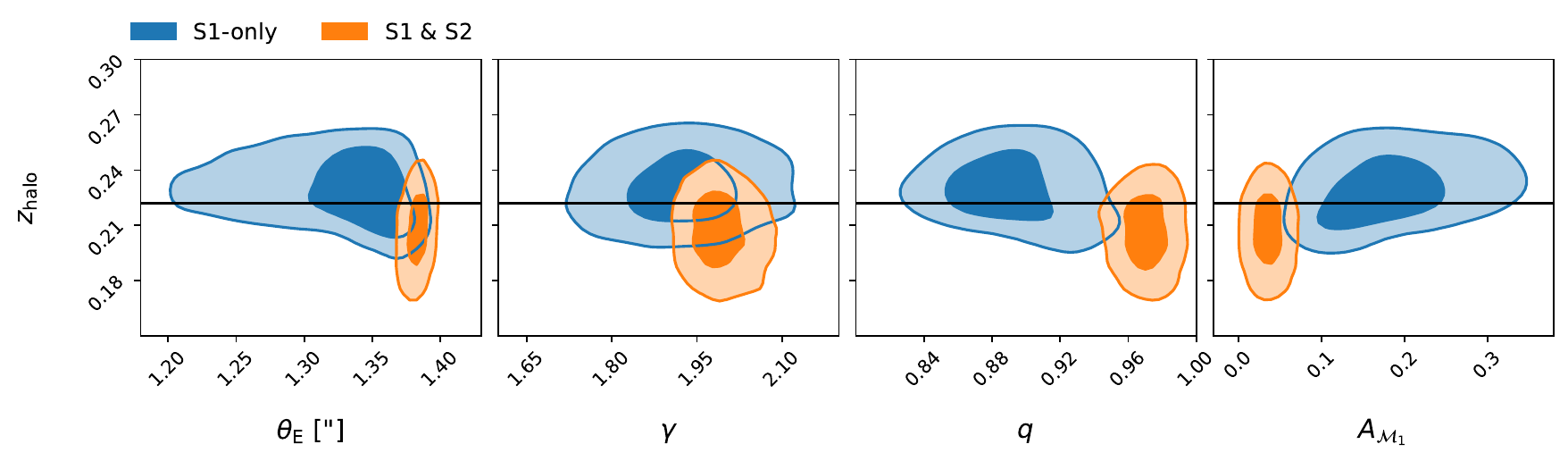}
    \includegraphics[width=\textwidth]{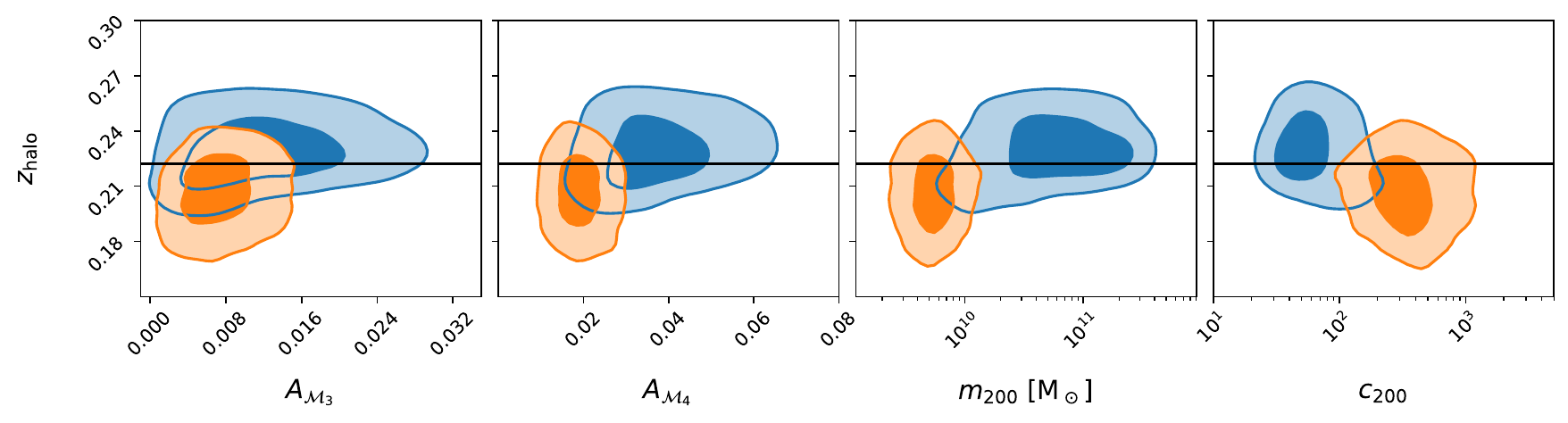}
    \includegraphics[width=\textwidth]{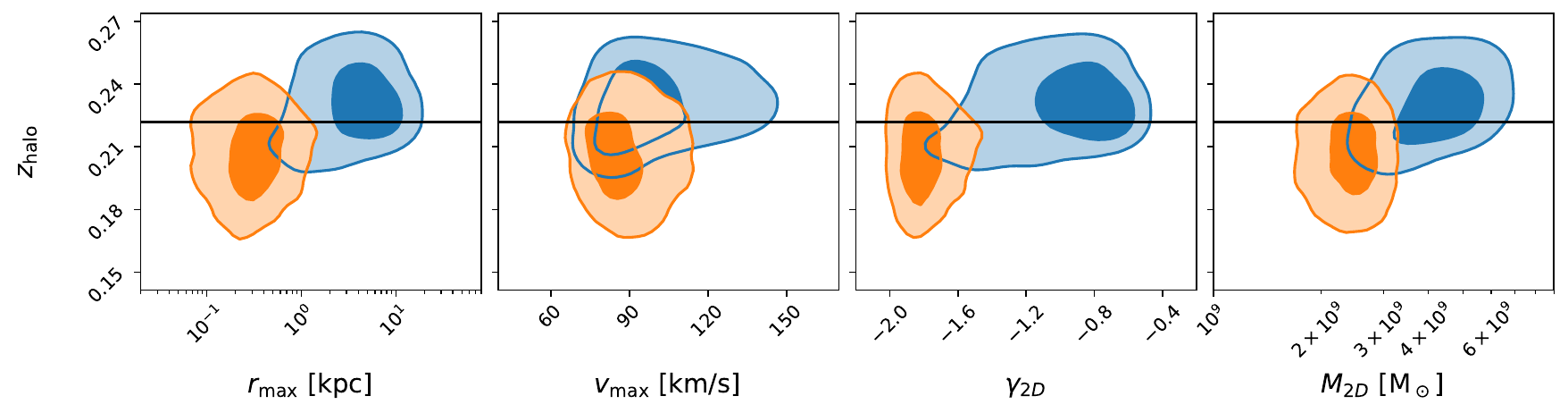}
    \caption{ Projected 2D posteriors that show the (lack of) correlation between the redshift and other model parameters. 
 The black line splits samples into those in front or behind the main deflector. The inferred parameters change only slightly with the redshift values allowed by our posterior.  }
    \label{fig:redshift_posterior_degeneracies}
\end{figure*}

Consequently, our inferred  parameters do not differ much from those found in previous studies that assume it to be a subhalo. The top right panel of Figure \ref{fig:triangle_main_cm-relation} shows the posterior of mass and concentration of the halo, which we constrain to $\log_{10} m_{200} / $M$_\odot = {9.72}^{+0.17}_{-0.18}$ and $\log_{10} c_{200} = {2.55}^{+0.31}_{-0.25}$.  The inferred perturber would be a $\sim 4 \sigma$ outlier as a field halo assuming the mass-concentration relations expected in $\Lambda$CDM extrapolated to small mass haloes.\footnote{\citet[][]{2024arXiv240901758S} consider haloes down to a mass of $10^{9.5} $M$_\odot$, therefore requiring only minimal extrapolation for comparison.}  We also find that the parameters of the subhalo and main deflector do not change much with the inferred redshift value (see Figure \ref{fig:redshift_posterior_degeneracies}) considering both their mean and standard deviations.
 
\begin{figure}
    \centering
    \includegraphics[width=\columnwidth]{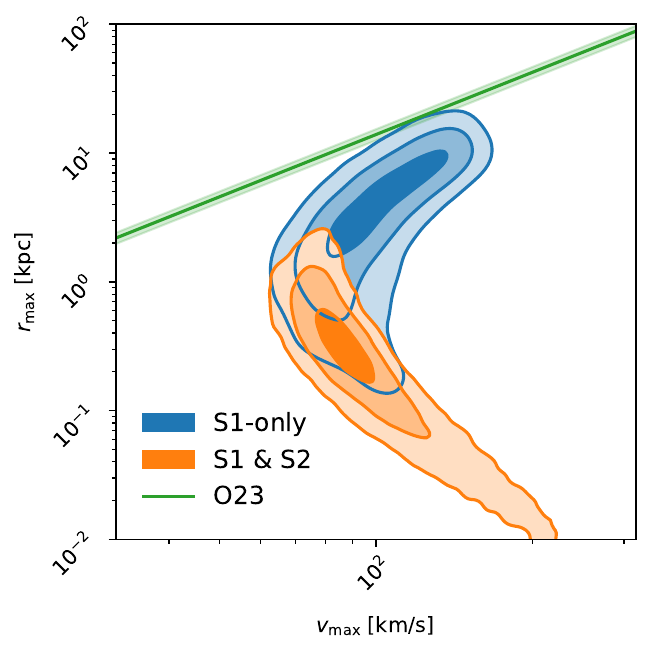}
    \caption{The first three $\sigma$ contours of our posteriors on $v_{\rm max}$ and $r_{\rm max}$ in blue and orange. Green shows the  best fit $v_{\rm max}$-$r_{\rm max}$ relation derived by \citet[][O23]{2023MNRAS.521.2342O} using the ShinUchuu simulation \citep[][]{2020MNRAS.492.3662I,2023MNRAS.518..157M}. The light green shows the $2\sigma$ error of this relation ($\sim10$ percent). 
    }
    \label{fig:rmax_vmax_posterior}
\end{figure}

If the dark halo is a subhalo as expected, its properties are better characterized by relations between the maximum circular velocity, $v_{\rm max}$, and  the corresponding radius, $r_{\rm max}$ \citep[see e.g.][]{2017MNRAS.466.4974M,2023MNRAS.518..157M, 2024MNRAS.528.1757O}.
We derive from our posterior that $v_{\rm max} = {87.85}^{+16.86}_{-9.81} $ km/s  and
$r_{\rm max} = {0.28}^{+0.27}_{-0.16}$ kpc with the inclusion of S2 (and $v_{\rm max} = {97.2}^{+25.7}_{-15.1} $ km/s and $r_{\rm max} = {3.49}^{+4.24}_{-2.19}$ kpc without S2).
Using the relation presented by \citet[][]{2023MNRAS.521.2342O}, which was based upon the ShinUchuu simulation \citep[see][]{2020MNRAS.492.3662I,2023MNRAS.518..157M}, we find again that the subhalo is a $> 5\sigma$ outlier (although only $3 \sigma$ when S2 is not included, see Figure~\ref{fig:rmax_vmax_posterior}). We note that the distribution is highly skewed towards it being more of an outlier, making a concrete statement on the level of this tension difficult due to a lack of samples close to the relation. Furthermore, the relation we use for comparison does not account for any redshift dependence.

Next, we compare our results with  $\Lambda$CDM expectations from the Illustris TNG50 simulations \citep[][]{2018MNRAS.473.4077P,2019ComAC...6....2N}. 
Following \citet[][]{2021MNRAS.507.1662M}, \citetalias[][]{2024arXiv240712910D}, and \citetalias[][]{2024arXiv240811090M}, we calculate  the average slope of the dark halo (within 0.75 to 1.25 kpc), $\gamma_{2D}$, and the mass it contains in projection up to 1 kpc,  $ M_{\rm 2D}$. With $\log_{10} M_{\rm 2D} / $M$_\odot = {9.61}^{+0.11}_{-0.12}$ and  $\gamma_{\rm 2D} = {-1.00}^{+0.28}_{-0.40}$ our S1-only model would also fall within the bulk of the  distribution of analogous simulated haloes and not necessarily be an outlier regarding its concentration \citepalias[see][]{2024arXiv240712910D}.  However, including the second source, we find that $\log_{10} M_{\rm 2D} / $M$_\odot = {9.38}^{+0.06}_{-0.08}$ and  $\gamma_{\rm 2D} = {-1.81}^{+0.15}_{-0.11}$, such that the subhalo is a significant outlier. This result is in  agreement with \citetalias{2024arXiv240811090M}, and shows the power of a compound lens for breaking modelling degeneracies.

\begin{figure}
    \centering
    \includegraphics[width=\columnwidth]{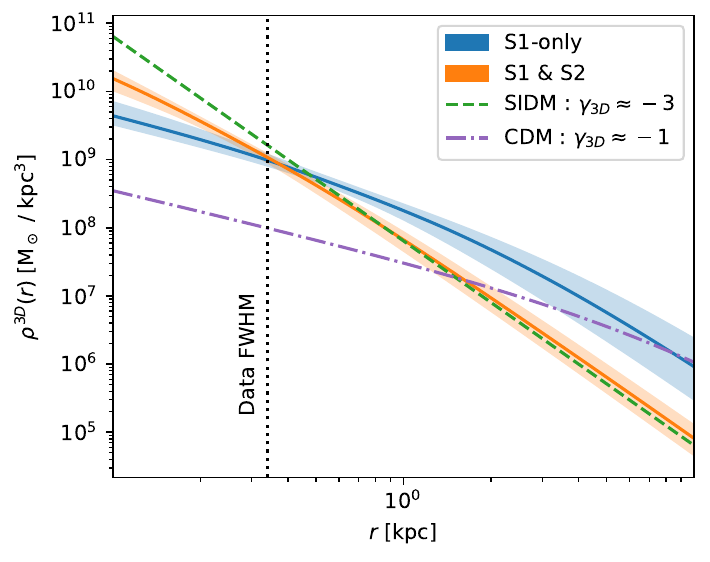}
    \caption{The reconstructed dark halo profiles for our S1-only and S1\&S2 models. For comparison we show a power law with a slope of $\gamma\approx-3$ as expected for an SIDM halo \citep[see e.g.][]{2021MNRAS.505.5327T}. We also show the expected density profile of a CDM halo with equivalent mass;  the central slope should be $\gamma \approx -1$.
    The dotted vertical line shows the size scale that corresponds to the FWHM of the PSF (of the F814W HST data). }
    \label{fig:profile_comparison}
\end{figure}

Figure \ref{fig:profile_comparison} shows our posterior constraints on the three-dimensional density profile of the dark halo. As seen previously, our S1\&S2 model finds a much steeper profile than the one  expected for an equally massive CDM halo, for which we expect $\gamma_{3D} \approx -1$ in the centre. However, with a  $\gamma_{3D} \approx -3$ , our density profile is in good agreement with the expectations from an SIDM core collapse scenario \citep[][]{2021MNRAS.505.5327T}. Deviations from this power law appear only below the resolved scales, i.e. scales smaller than the full width at half maximum (FWHM) of the PSF of our HST data. This result highlights the necessity of high-resolution data to obtain stronger constraints on the density profile of this subhalo in the future. Our S1-only model shows a slope more compatible with CDM expectations, but this is an artifact of the degeneracy between substructure parameters and lens multipoles for the single plane dataset.

While the inclusion of compound lensing did not improve constraints on the redshift of the perturber, it does affect the subhalo parameters. Without the additional constraints from the second source, it is possible to have a subhalo with a larger $m_{\rm 200}$ if the multipole perturbations adjust accordingly.  Figures \ref{fig:s1_only_max_like_model} and \ref{fig:s1_s2_max_like_model} show that this will overall keep the convergence similar.

One novel feature of our work is that we simultaneously model the lens light and the dark halo. Doing this improves our constraints on its redshift, because it rules out high mass, low redshift haloes that would also lens the main deflector light. Figure \ref{fig:redshift_posterior} shows that without this effect we would have inferred a slightly lower redshift but still compatible with a subhalo.

We further obtain the uncertainty maps $\sigma_{\vec \kappa}$ of the mass distribution. We can, therefore, also see which parts of the mass distribution are the least constrained. We find higher uncertainties near the center of the main deflector and the dark halo, pointing either towards a deficiency of the chosen mass profile (of the subhalo) or the requirement of more data to constrain these parts of the mass distribution. This could be achieved, for example, by the inclusion of Muse data of the third source \citep[][]{2020MNRAS.497.1654C} or the inclusion of additional data from HST or JWST imaging in various bands \citepalias[see e.g. ][]{2024MNRAS.528.7564B}.

In contrast to previous works, we do not find that the truncation radius $r_t$ is well constrained by the data. The reason for this could be the free redshift, the inclusion of the $\mathcal{M}_1$ multipole, or a combination of the two.
With our chosen range with find that 95 percent of posterior samples fall  above $r_t = 3.43 $ \arcsec, which is larger than the size of the data cutout we analyse in this work. We further find that the parameters $r_s$ and $r_t$ of the perturber are (almost) interchangable, as  the smaller one mostly decides at which point the logarithmic slope of the dark halo changes from $-1$ to $-3$, and the larger one where it changes from $-3$ to $-5$. One of our chains in the first source only reconstruction found a good solution with an unconstrained $r_s$ while $r_t$ is constrained. As pointed out by \citet[][]{2009JCAP...01..015B}, a steeper cutoff might be more physically motivated for $r_t < r_s$. We therefore do not include this chain here, even though its inclusion would not significantly change our results. 

\subsection{Source Reconstruction}

Similar to \citet[][]{2024arXiv240608484G}, we find that the GP approach can struggle with high dynamic range in the source distribution. In particular, we find that some of the star-forming regions that are present in the first source \jackpot appear oversmoothed in our reconstruction. In Figures \ref{fig:s1_only_max_like_model} and \ref{fig:s1_s2_max_like_model}, we show not only the source reconstructions, $\vec s_1$ and  $\vec s_2$, but also the uncertainty maps,  $\sigma_{\vec s_1}$ and $\sigma_{\vec s_1}$. Looking into the uncertainty maps, we identify multiple smaller regions of higher uncertainty. These regions could indicate how the source model struggles with the  high dynamic range of features in the source, such as star-forming regions. Alternatively, this uncertainty could reflect some deficiencies in our assumed mass model. In a follow-up paper, we will consider the latter option by including convergence corrections in our reconstruction. To highlight the features we reconstruct with high significance, we further show the ratio of mean and standard deviations for both sources.

Previous studies used different regularization orders of the background sources \citep[see e.g.][ who used curvature and gradient, and SNR weighted regularizations]{2010MNRAS.408.1969V,2024MNRAS.528.7564B,2024arXiv240712910D,2024arXiv240811090M}. The GP approach to modelling the pixelated source allows us to be more agnostic about the correlation structure of the source or the potential corrections. In agreement with \citetalias[][]{2024MNRAS.528.7564B}, we find that the first source prefers a regularization that is close to gradient regularization. Meanwhile, the second source prefers a stronger regularization on smaller scales.

\section{Conclusions}
\label{sec:Conclusion}

In this work, we analysed the lens system \jackpot assuming an EPL+Shear model with multipoles and a parametric dark halo while allowing for the first time for a free redshift of the latter. Using an SVI-preconditioned HMC-within-Gibbs approach, we generate samples of posterior models that explain the data to the noise level. Our method entirely follows a forward modelling philosophy. We then discussed our results in the context of the tension that this subhalo poses for $\Lambda$CDM predictions. We summarise the conclusions of our analyses as follows:

First, the redshift of the dark structure in \jackpot can be constrained from its lensing effect on both the light of the main deflector and the lensed sources behind it. We find $z_{\rm halo}={0.207}^{+0.019}_{-0.019}$, which is consistent with the main deflector redshift. 
A model with a subhalo is preferred with a Bayes factor of $\log \alpha_{\rm Bayes} \approx 7$ over the field halo model. This value corresponds to posterior chance of 1 in 1000 that the halo is a field halo. The inferred values of $v_{\rm max} ={87.85}^{+16.86}_{-9.81} $ km/s and $r_{\rm max} = {0.28}^{+0.27}_{-0.16}$ kpc are in $>5\sigma$ tension with the predictions from $\Lambda$CDM. 
If, however, the dark halo is a field halo, it would still be an $\sim 4 \sigma$ outlier in $\Lambda$CDM given its mass, $\log_{10} m_{200} / $M$_\odot = {9.72}^{+0.17}_{-0.18}$, and concentration, $\log_{10} c_{200} = {2.55}^{+0.31}_{-0.25}$. 

Second, compound lensing does not render the redshift constraints  of the dark halo significantly more certain. However, it does help to break important degeneracies in the multipole perturbations of the main deflector. This is likely due to a larger radial coverage of the light probing the mass profile. In particular, we would expect multiple background sources at the same redshift to provide similar additional constraints on the multipoles. The additional constraints on the multipoles affect the inferred parameters describing the halo profile, e.g. its mass and concentration, but not significantly its redshift. Further degeneracies in the main deflector mass model could potentially be broken by including the kinematics of this lens \citep[see e.g.][]{2024MNRAS.528.3559T, 2024arXiv240811090M} or additional data from other bands \citepalias[see e.g.][]{2024MNRAS.528.7564B}.

Third, the tension that this subhalo shows with respect to $\Lambda$CDM predictions deserves more attention, as it might provide a smoking gun for dark matter self-interaction. The high concentration could be explained by a core undergoing gravothermal collapse \citep[see e.g.][]{2021MNRAS.507.1662M,2021MNRAS.505.5327T,2024JCAP...02..032Y}, but the existing HST data lacks the SNR or angular resolution to make such claims conclusive. In agreement with \citetalias{2024arXiv240811090M}, we find a power law density profile slope of $\gamma_{\rm 2D} = {-1.81}^{+0.15}_{-0.11}$ . This value is close to the $\gamma_{\rm 2D} \approx -2$ expected for gravothermally collapsed SIDM haloes \citep[see e.g.][]{2021MNRAS.505.5327T}. 

 The subhalo in \jackpot is very much in tension with $\Lambda$CDM. In the coming years, Euclid is expected to discover $\mathcal{O}(10^5)$ new lenses \citep{2015ApJ...811...20C}  and be sensitive to 2500 subhaloes \citep{2023MNRAS.521.2342O}. If the subhalo in \jackpot turns out to be typical, then it seems likely that CDM will soon be definitively ruled out and dark matter self interactions preferred.

\clearpage

\section*{Acknowledgements}
We thank Aymeric Galan, Giulia Despali, Quinn Minor, Konstantin Leyde, Dan Ryczanowski, and Andrew Gow, for the insightful discussions on several aspects of this paper. This project
has received funding from the European Research Council
(ERC) under the European Union’s Horizon 2020 research
and innovation programme (LensEra: grant agreement No
945536). TEC is funded by the Royal Society through a University Research Fellowship. Numerical computations were done on the Sciama High Performance Compute (HPC) cluster which is supported by the ICG, SEPNet and the University of Portsmouth. For the purpose of open access, the authors have applied a Creative Commons Attribution (CC BY) licence to any Author Accepted Manuscript version arising.

%%%%%%%%%%%%%%%%%%%%%%%%%%%%%%%%%%%%%%%%%%%%%%%%%%
\section*{Data Availability}

Supporting research data are available on reasonable request from the
corresponding author and from the HST archive.



\bibliographystyle{mnras}
\bibliography{paper}

%\begin{thebibliography}{99}
%\end{thebibliography}

%%%%%%%%%%%%%%%%%%%%%%%%%%%%%%%%%%%%%%%%%%%%%%%%%%

%%%%%%%%%%%%%%%%% APPENDICES %%%%%%%%%%%%%%%%%%%%%

\appendix

% \section{Some extra material}

% If you want to present additional material which would interrupt the flow of the main paper,
% it can be placed in an Appendix which appears after the list of references.

\renewcommand{\arraystretch}{1.3}

% table contains main posterior
\begin{table*}

	\centering
	\caption{This table presents the priors and posterior values of all our model parameters after the HMC stage. We refer to uniform distributions as $\mathcal{U}$, normal distributions as $\mathcal{G}$, halfnormal distributions as $\mathcal{G}_+$, and  the linear distribution on an interval as described in appendix \ref{sec:linear_distribution} as $\mathcal{T}$. The agreement describes within how many standard deviations of this posterior, our reported medians match those of previous studies (rounded to the next higher integer).  }
	\label{tab:HMCgibbsMain}
\small
\begin{tabular}{lcccccc} 
    \hline
   & \textbf{Parameter} & \textbf{Prior}  & \multicolumn{2}{c}{\textbf{Posterior} (Median $\pm 1\sigma$)}  & \textbf{Agreement within}    \\
    & & & Single Plane & Double Plane  & Author : $N\sigma$ \\
    \hline
 \hline
$\vec s_{\rm lens}$ &$A $ & $10^{\mathcal{U}(-5,4)}$  & ${1.9}^{+3931}_{-1.9}$ & ${1.3}^{+3477}_{-1.3}$ \\
&$ e_x $ & $\mathcal{G}(0,0.1)$  & ${0.001}^{+0.1}_{-0.1}$ & ${0.000}^{+0.1}_{-0.1}$  \\
&$ e_y $ & $\mathcal{G}(0,0.1)$  & ${0.001}^{+0.099}_{-0.100}$ & ${-0.001}^{+0.099}_{-0.099}$  \\
&$ {\sigma} [\arcsec]$ & Section \ref{sec:ParametricLightModel}  & ${0.012}^{+0.001}_{-0.001}$ & ${0.012}^{+0.001}_{-0.001}$  \\
&$x [\arcsec]$ & $\mathcal{G}(0,0.1)$  & ${-0.007}^{+0.096}_{-0.13}$ & ${-0.004}^{+0.093}_{-0.13}$  \\
&$y [\arcsec]$ & $\mathcal{G}(0,0.1)$  & ${0.017}^{+0.11}_{-0.11}$ & ${0.015}^{+0.11}_{-0.11}$   \\
 \hline
$\vec \lambda_{s1}$ & $n$ & $\mathcal{T}(-1, 0.1, 10)$  & ${0.201}^{+0.094}_{-0.064}$ & ${0.170}^{+0.064}_{-0.047}$  \\
& $ \zeta$ & $e^{\mathcal{G}(2.1,1.1)}$  & ${27.6}^{+28.6}_{-11.6}$ & ${27.3}^{+28.7}_{-11.6}$  \\
& $\sigma$ & $10^{\mathcal{U}(-5,5)}$  & ${0.214}^{+0.034}_{-0.025}$ & ${0.217}^{+0.029}_{-0.022}$  \\
 \hline
$\vec \lambda_{s2}$& $n$ & $\mathcal{T}(-1, 0.1, 10)$  & - & ${4.0}^{+4.2}_{-3.3}$  \\
& $\zeta$ & $e^{\mathcal{G}(2.1,1.1)}$  & - & ${4.59}^{+1.38}_{-0.87}$  \\
& $\sigma$ & $10^{\mathcal{U}(-5,5)}$  & - & ${0.027}^{+0.004}_{-0.003}$  \\
 \hline
$\vec \eta_{\rm EPL, lens} $ &$\theta_E[\arcsec]$ & $\mathcal{U}(1.2,1.6)$  & ${1.330}^{+0.035}_{-0.067}$ & ${1.383}^{+0.005}_{-0.006}$  & \ballard:2, \minor:2\\
& $e_x$ & $\mathcal{G}(0,0.1)$  & ${0.022}^{+0.010}_{-0.009}$ & ${0.002}^{+0.004}_{-0.004}$    \\
& $e_y$ & $\mathcal{G}(0,0.1)$  & ${0.051}^{+0.018}_{-0.018}$ & ${0.014}^{+0.007}_{-0.007}$  \\
 & $\phi$ & -   & ${0.579}^{+0.080}_{-0.082}$ & ${0.71}^{+0.15}_{-0.17}$  & \ballard:2, \minor:2\\
 & $q$ & -   & ${0.893}^{+0.032}_{-0.032}$ & ${0.971}^{+0.012}_{-0.012}$  & \ballard:1, \minor:1\\
& $\gamma$ & $\mathcal{G}(2.0,0.2)$  & ${1.928}^{+0.100}_{-0.098}$ & ${1.990}^{+0.049}_{-0.041}$  & \ballard:2, \minor:2 \\
& $x [\arcsec]$ & $\mathcal{G}(0.0,0.5)$  & ${-0.011}^{+0.022}_{-0.023}$ & ${0.050}^{+0.009}_{-0.009}$  & \ballard:2, \minor:1 \\
& $y [\arcsec]$ & $\mathcal{G}(0.0,0.5)$  & ${0.104}^{+0.029}_{-0.025}$ & ${0.053}^{+0.013}_{-0.012}$ & \ballard:2, \minor:2 \\
\hline $\vec \eta_{\Gamma}$& $\Gamma_x$ & $\mathcal{U}(-0.2,0.2)$  & ${0.057}^{+0.007}_{-0.007}$ & ${0.051}^{+0.003}_{-0.003}$ & \ballard:4, \minor:10 \\
 & $\Gamma_y$ & $\mathcal{U}(-0.2,0.2)$  & ${-0.021}^{+0.008}_{-0.008}$ & ${-0.039}^{+0.003}_{-0.004}$ & \ballard:4 , \minor:6\\
\hline $\eta_{\mathcal{M}_1}$ & $e_x$ & $\mathcal{G}(0,0.1)$  & ${0.062}^{+0.037}_{-0.026}$ & ${-0.003}^{+0.006}_{-0.006}$   \\
 & $e_y$ & $\mathcal{G}(0,0.1)$  & ${-0.076}^{+0.030}_{-0.045}$ & ${-0.017}^{+0.009}_{-0.009}$  \\
 & $A_{\mathcal{M}_1}$ & -  & ${0.180}^{+0.087}_{-0.067}$ & ${0.036}^{+0.016}_{-0.016}$  \\
 & $ \phi_{\mathcal{M}_1}$ & -  & ${-0.91}^{+0.14}_{-0.12}$ & ${-1.71}^{+0.43}_{-0.37}$  \\
\hline $\eta_{\mathcal{M}_3}$ & $e_x$ & $\mathcal{G}(0,0.1)$  & ${-0.006}^{+0.004}_{-0.005}$ & ${0.001}^{+0.002}_{-0.002}$  \\
& $e_y$ & $\mathcal{G}(0,0.1)$  & ${-0.001}^{+0.002}_{-0.002}$ & ${0.003}^{+0.002}_{-0.002}$  \\
 & $A_{\mathcal{M}_3}$ & -  & ${0.012}^{+0.009}_{-0.007}$ & ${0.008}^{+0.004}_{-0.004}$  & \ballard:3, \minor:2 \\
 & $ \phi_{\mathcal{M}_3}$ & -  & ${-0.87}^{+1.78}_{-0.11}$ & ${0.39}^{+0.18}_{-0.13}$  \\
\hline $\eta_{\mathcal{M}_4}$ & $e_x$ & $\mathcal{G}(0,0.1)$  & ${0.013}^{+0.006}_{-0.004}$ & ${0.006}^{+0.002}_{-0.002}$    \\
& $e_y$ & $\mathcal{G}(0,0.1)$  & ${-0.013}^{+0.004}_{-0.005}$ & ${-0.007}^{+0.002}_{-0.002}$  \\
 & $A_{\mathcal{M}_4}$ & -  & ${0.037}^{+0.014}_{-0.011}$ & ${0.019}^{+0.005}_{-0.005}$  & \ballard:2, \minor:1\\
 & $ \phi_{\mathcal{M}_4}$ & -  & ${-0.199}^{+0.030}_{-0.029}$ & ${-0.216}^{+0.043}_{-0.040}$  \\
\hline $\vec \eta_{\rm EPL, s1}$ & $\theta_E [\arcsec]$ & $\mathcal{U}(0,1.0)$  & - & ${0.152}^{+0.038}_{-0.032}$  & \\
&$e_x$ & $\mathcal{G}(0,0.1)$  & - & ${0.094}^{+0.055}_{-0.050}$  \\
& $e_y$ & $\mathcal{G}(0,0.1)$  & - & ${0.109}^{+0.058}_{-0.054}$  \\
 & $\phi$ & -   & - & ${0.43}^{+0.21}_{-0.22}$ & \\
 & $q$ & -   & - & ${0.731}^{+0.061}_{-0.070}$  \\
  \hline
& $\sigma_{\rm bg}$ & $\mathcal{G}_+(0.0128)$  & ${0.013}^{+0.000}_{-0.000}$ & ${0.013}^{+0.000}_{-0.000}$  \\
 \hline
        \end{tabular}
       
\end{table*}

\begin{table*}

	\centering
	\caption{Same as Table \ref{tab:HMCgibbsMain} but for the halo Parameters. The halos center coordinates are defined relative to $x_0=-0.68$, $y_0=1.0$ (or where this position is mapped to, if the halo is behind the main deflector.). We note that the compared references make different assumptions assumptions in their modelling approaches, including on the shape of the halo profile, the amount of supersampling, and the inclusion of the second source, therefore explaining some discrepancies. We also compare the halo mass with \citet[][N24]{2024MNRAS.52710480N}.  }
	\label{tab:HMCgibbsSub}
 \small
\begin{tabular}{lccccc} 
    \hline
 & \textbf{Parameter} & \textbf{Prior}  & \multicolumn{2}{c}{\textbf{Posterior} (Median $\pm 1\sigma$)}  &  \textbf{Agreement within}   \\
    & & & Single Plane & Double Plane &  Author : $N\sigma$ \\
    \hline
 
 \hline
$\vec \eta_{\rm tNFW}$ & $\log_{10} k_s $ & ${\mathcal{U}(-10,15)}$  & ${-0.70}^{+0.29}_{-0.16}$ & ${0.30}^{+0.52}_{-0.39}$  \\
& $\log_{10} r_s / \arcsec$ & ${\mathcal{U}(-10,15)}$  & ${-0.38}^{+0.35}_{-0.42}$ & ${-1.42}^{+0.29}_{-0.37}$  \\
& $\log_{10} r_t / \arcsec$ & $\mathcal{U}(-10,15)$  & ${7.7}^{+4.9}_{-5.1}$ & ${7.4}^{+5.2}_{-5.2}$  \\
& $x [\arcsec]$ & $\mathcal{G}(-0.3,0.3)$  & ${0.012}^{+0.020}_{-0.023}$ & ${0.047}^{+0.018}_{-0.018}$  \\
& $y [\arcsec]$ & $\mathcal{G}(-0.3,0.3)$  & ${-0.004}^{+0.028}_{-0.027}$ & ${-0.029}^{+0.025}_{-0.025}$  \\
 & z & $\mathcal{U}(0.001,0.608)$  & ${0.229}^{+0.018}_{-0.015}$ & ${0.207}^{+0.019}_{-0.019}$  \\
 & $\log_{10} M_{\rm 2D} / $M$_\odot$ & -  & ${9.61}^{+0.11}_{-0.12}$ & ${9.382}^{+0.064}_{-0.080}$ & \minor:1 \\
 & $\gamma_{\rm 2D}$ & -  & ${-1.00}^{+0.28}_{-0.40}$ & ${-1.81}^{+0.15}_{-0.11}$  & \minor:2 \\
 & $\log_{10} m_{200} / $M$_\odot $ & -  & ${10.69}^{+0.50}_{-0.49}$ & ${9.72}^{+0.17}_{-0.18}$  &  \ballard:5, \minor:1, \despali:4 \\
 & $\log_{10} c_{200}$ &-   & ${1.77}^{+0.27}_{-0.18}$ & ${2.55}^{+0.31}_{-0.25}$ & \ballard:3, \minor:1, \despali:2\\
 & $\log_{10} M_{\rm tot}  / $M$_\odot$ & -  & ${11.52}^{+0.61}_{-0.66}$ & ${10.45}^{+0.32}_{-0.47}$ & \ballard:2 , \minor:2, \nightingale:5 \\

 & $ v_{\rm max}$ [km/s]  & -  & ${97.2}^{+25.7}_{-15.1}$ & ${87.85}^{+16.86}_{-9.81}$ &  \\
 & $ r_{\rm max}$ [kpc]  & -  & ${3.49}^{+4.24}_{-2.19}$ & ${0.28}^{+0.27}_{-0.16}$ \\
 
        \hline
        \end{tabular}
\end{table*}

\section{Jeffreys prior}
\label{sec:linear_distribution}
We employ approximations to Jeffreys prior \citep[][]{1946RSPSA.186..453J} for the parameters that enter the Matern power spectrum.
For the slope parameter $n$ we find that this prior roughly follows a truncated linear distribution. We define this distribution, $\mathcal{T}$, on the interval $[x_{\rm min},x_{\rm max}]$ via position $x_0$, at which it intercepts the x-axis. The resulting Probability distribution is:
\begin{equation}
    \mathcal{T}(x,x_0,a,b) = \frac{2(x-x_0)}{(x_{\rm max}-x_0)^2 - (x_{\rm min}-x_0)^2} \,.
\end{equation}
We found that this is a good approximation to Jeffreys prior for the index $n$ of the Matern kernel (within the ranges of $n$ we consider here).
For the parameter $\zeta$, we find that Jeffreys prior is well approximated by a log-normal distribution, while the standard deviation of this power spectrum, $\sigma$, follows approximately a log-Uniform distribution. The explicit parameters that enter these distributions are shown in Table \ref{tab:HMCgibbsMain} and only depend very little on the number of pixels of the field (if there are more than a few).

%%%%%%%%%%%%%%%%%%%%%%%%%%%%%%%%%%%%%%%%%%%%%%%%%%

% Don't change these lines
\bsp	% typesetting comment
\label{lastpage}
\end{document}